\documentclass[runningheads]{llncs}

\makeatletter
\@ifundefined{lhs2tex.lhs2tex.sty.read}%
  {\@namedef{lhs2tex.lhs2tex.sty.read}{}%
   \newcommand\SkipToFmtEnd{}%
   \newcommand\EndFmtInput{}%
   \long\def\SkipToFmtEnd#1\EndFmtInput{}%
  }\SkipToFmtEnd

\newcommand\ReadOnlyOnce[1]{\@ifundefined{#1}{\@namedef{#1}{}}\SkipToFmtEnd}
\usepackage{amstext}
\usepackage{amssymb}
\usepackage{stmaryrd}
\DeclareFontFamily{OT1}{cmtex}{}
\DeclareFontShape{OT1}{cmtex}{m}{n}
  {<5><6><7><8>cmtex8
   <9>cmtex9
   <10><10.95><12><14.4><17.28><20.74><24.88>cmtex10}{}
\DeclareFontShape{OT1}{cmtex}{m}{it}
  {<-> ssub * cmtt/m/it}{}

\DeclareFontShape{OT1}{cmtt}{bx}{n}
  {<5><6><7><8>cmtt8
   <9>cmbtt9
   <10><10.95><12><14.4><17.28><20.74><24.88>cmbtt10}{}
\DeclareFontShape{OT1}{cmtex}{bx}{n}
  {<-> ssub * cmtt/bx/n}{}

\newcommand{\Conid}[1]{\mathit{#1}}
\newcommand{\Varid}[1]{\mathit{#1}}
\newcommand{\anonymous}{\kern0.06em \vbox{\hrule\@width.5em}}
\newcommand{\plus}{\mathbin{+\!\!\!+}}
\newcommand{\bind}{\mathbin{>\!\!\!>\mkern-6.7mu=}}

\usepackage{polytable}

\@ifundefined{mathindent}%
  {\newdimen\mathindent\mathindent\leftmargini}%
  {}%

\def\resethooks{%
  \global\let\SaveRestoreHook\empty
  \global\let\ColumnHook\empty}
\newcommand*{\savecolumns}[1][default]%
  {\g@addto@macro\SaveRestoreHook{\savecolumns[#1]}}
\newcommand*{\restorecolumns}[1][default]%
  {\g@addto@macro\SaveRestoreHook{\restorecolumns[#1]}}
\newcommand*{\aligncolumn}[2]%
  {\g@addto@macro\ColumnHook{\column{#1}{#2}}}

\resethooks

\newcommand{\onelinecommentchars}{\quad-{}- }
\newcommand{\commentbeginchars}{\enskip\{-}
\newcommand{\commentendchars}{-\}\enskip}

\newcommand{\visiblecomments}{%
  \let\onelinecomment=\onelinecommentchars
  \let\commentbegin=\commentbeginchars
  \let\commentend=\commentendchars}

\newcommand{\invisiblecomments}{%
  \let\onelinecomment=\empty
  \let\commentbegin=\empty
  \let\commentend=\empty}

\visiblecomments

\newlength{\blanklineskip}
\newcommand{\hsindent}[1]{\quad}
\let\hspre\empty
\let\hspost\empty

\EndFmtInput
\makeatother
\ReadOnlyOnce{polycode.fmt}%
\makeatletter

\newcommand{\hsnewpar}[1]%
  {{\parskip=0pt\parindent=0pt\par\vskip #1\noindent}}

\newcommand{\hscodestyle}{}

\newcommand{\sethscode}[1]%
  {\expandafter\let\expandafter\hscode\csname #1\endcsname
   \expandafter\let\expandafter\endhscode\csname end#1\endcsname}

  {\par\noindent
   \advance\leftskip\mathindent
   \hscodestyle
   \let\\=\@normalcr
   \let\hspre\(\let\hspost\)%
   \pboxed}%
  {\endpboxed\)%
   \par\noindent
   \ignorespacesafterend}

  {\hsnewpar\abovedisplayskip
   \advance\leftskip\mathindent
   \hscodestyle
   \let\hspre\(\let\hspost\)%
   \pboxed}%
  {\endpboxed%
   \hsnewpar\belowdisplayskip
   \ignorespacesafterend}

  {\hsnewpar\abovedisplayskip
   \advance\leftskip\mathindent
   \hscodestyle
   \let\\=\@normalcr
   \(\pboxed}%
  {\endpboxed\)%
   \hsnewpar\belowdisplayskip
   \ignorespacesafterend}

\newcommand{\plainhs}{\sethscode{plainhscode}}

\plainhs

  {\hsnewpar\abovedisplayskip
   \advance\leftskip\mathindent
   \hscodestyle
   \let\\=\@normalcr
   \(\parray}%
  {\endparray\)%
   \hsnewpar\belowdisplayskip
   \ignorespacesafterend}

  {\parray}{\endparray}

  {\(\parray}{\endparray\)}

\def\codeframewidth{\arrayrulewidth}
\RequirePackage{calc}

  {\parskip=\abovedisplayskip\par\noindent
   \hscodestyle
   \arrayrulewidth=\codeframewidth
   \tabular{@{}|p{\linewidth-2\arraycolsep-2\arrayrulewidth-2pt}|@{}}%
   \hline\framedhslinecorrect\\{-1.5ex}%
   \let\endoflinesave=\\
   \let\\=\@normalcr
   \(\pboxed}%
  {\endpboxed\)%
   \framedhslinecorrect\endoflinesave{.5ex}\hline
   \endtabular
   \parskip=\belowdisplayskip\par\noindent
   \ignorespacesafterend}

\newcommand{\framedhslinecorrect}[2]%
  {#1[#2]}

  {\(\def\column##1##2{}%
   \let\>\undefined\let\<\undefined\let\\\undefined
   \newcommand\>[1][]{}\newcommand\<[1][]{}\newcommand\\[1][]{}%
   \def\fromto##1##2##3{##3}%
   }{\) }%

  {\let\orighscode=\hscode
   \let\origendhscode=\endhscode
   \def\endhscode{\def\hscode{\endgroup\def\@currenvir{hscode}\\}\begingroup}
   \orighscode\def\hscode{\endgroup\def\@currenvir{hscode}}}%
  {\origendhscode
   \global\let\hscode=\orighscode
   \global\let\endhscode=\origendhscode}%

\makeatother
\EndFmtInput
\usepackage{amsmath}
\usepackage{caption}
\usepackage{float}
\usepackage{xspace}
\usepackage{xcolor}
\usepackage{comment}
\usepackage{graphicx}
\input{haskell_newcolorsBW.sty}
\input{agBW.sty}

\renewcommand{\hscodestyle}{\small}

\begin{document}

\title{Zipping Strategies and Attribute Grammars}


\author{Jos\'e Nuno Macedo\inst{1}\orcidID{0000-0002-0282-5060} \and
Marcos Viera\inst{2} \and
Jo\~ao Saraiva\inst{1}\orcidID{0000-0002-5686-7151}}
\authorrunning{J. N. Macedo et al.}

\institute{University of Minho, Braga, Portugal\\ 
\email{jose.n.macedo@inesctec.pt} \\ 
\email{saraiva@di.uminho.pt} \\
\and Universidad de la Rep\'ublica, Montevideo, Uruguay\\
\email{mviera@fing.edu.uy}}

\maketitle

\begin{abstract}

Strategic term rewriting and attribute grammars are two powerful programming techniques widely used in language engineering. The former, relies on strategies to apply term rewrite rules in defining language transformations, while the latter is suitable to express context-dependent language processing algorithms. Each of these techniques, however, is usually implemented by its own powerful and large language processor system. As a result, it makes such systems harder to extend and to combine.

In this paper, we present the embedding of both strategic tree rewriting and attribute grammars in a zipper-based, purely functional setting. Zippers provide a simple, but generic tree-walk mechanism that is the building block technique we use to express the purely-functional embedding of both techniques. 
The embedding of the two techniques in the same setting has several advantages: First, we easily combine/zip attribute grammars and strategies, thus providing language engineers the best of the two worlds. Second, the combined embedding is easier to maintain and extend since it is written in a concise and uniform setting. This results in a very small library 
which is able to express advanced (static) analysis and transformation tasks. We show the expressive power of our library in optimizing \ensuremath{\Conid{Haskell}} let expressions, expressing several \ensuremath{\Conid{Haskell}} refactorings and solving several language processing tasks of the LDTA Tool Challenge.



%
%
%
%

\end{abstract}


\section{Introduction}
\label{sec1}

Since Algol was designed in the 60's, as the first high-level
programming language~\cite{algol68}, languages have evolved
dramatically. In fact, modern languages offer powerful syntactic and
semantic mechanisms that do improve programmers productivity. In
response to such developments, the software language engineering
community also developed advanced techniques to specify those new
language mechanisms.

Strategic term rewriting~\cite{strategies97} and Attribute Grammars
(AG)~\cite{knuth1968semantics} have a long history in supporting the
development of modern software language analysis, transformations and
optimizations. The former, relies on strategies to traverse a tree while
applying a set of rewrite rules, while the latter is suitable to
express context-dependent language processing algorithms. Many
language engineering systems have been developed supporting both
AGs~\cite{eli,syngen,lrc,lisa,jastadd,uuag,silver} and rewriting
strategies~\cite{asfsdf,tom,strafunski,txl,kiama,stratego}. These
powerful systems, however, are large systems supporting their own AG
or strategic specification language, thus requiring a considerable
developing effort to extend and combine.

A more flexible approach is obtained when we consider the embedding of
such techniques in a general purpose language. Language embeddings,
however, usually rely on advanced mechanisms of the host language,
which makes them difficult to combine. For example,
Strafunski~\cite{strafunski} offers a powerful embedding of strategic
term rewriting in \ensuremath{\Conid{Haskell}}, but it can not be easily combined with
the \ensuremath{\Conid{Haskell}} embedding of AGs as provided
in~\cite{Moor00firstclass,zipperAG}. The former works directly on the
underlying tree, while the later on a \textit{zipper} representation
of the tree.

In this paper, we present the embedding of both strategic tree
rewriting and attribute grammars in a zipper-based, purely functional
setting. Generic zippers~\cite{thezipper} is a simple, but generic
tree-walk mechanism to navigate both on homogeneous and heterogeneous
data structures. Traversals on heterogeneous data structures is the
main ingredient of both strategies and AGs. Thus, zippers provide the
building block mechanism we will reuse to express the
purely-functional embedding of both techniques. The embedding of the
two techniques in the same setting has several advantages: First, we
easily combine/zip attribute grammars and strategies, thus providing
language engineers the best of the two worlds. Second, the combined
embedding is easier to maintain and extend since it is written in a
concise and uniform setting. This results in a very small library (200
lines of \ensuremath{\Conid{Haskell}} code) which is able to express advanced (static)
analysis and transformation tasks.

The purpose of this paper is three-fold:

\begin{itemize}

\item Firstly, we present a simple, yet powerful embedding of
strategic term rewriting using generic zippers. This results in a
simple and concise library, named \ensuremath{\Conid{Ztrategic}}, that is easy to maintain and
update. Moreover, our embedding has the expressiveness of the
Strafunski library~\cite{strafunski}.

\item Secondly, this new strategic term rewriting embedding can
easily be combined with a zipper-based embedding of attribute
grammars~\cite{scp2016,memoAG19}. In fact, by relying on the same
generic tree-traversal mechanism, the zippers, (zipper-based)
strategies can access (zipper-based) AG functional definitions, and
vice versa. Such a joint embedding results in a multi-paradigm
embedding of the two language engineering techniques. We show two
examples of the expressive power of such embedding: First, we access
attribute values in strategies so that we express non-trivial
context-dependent tree rewriting. Second, strategies are used to
define \textit{attribute propagation patterns}~\cite{eli,uuag}, which
are widely used to eliminate (polluting) copy rules from AG
specifications.

\item Thirdly, we apply \ensuremath{\Conid{Ztrategic}} in real language engineering problems, namely, in optimizing
\ensuremath{\Conid{Haskell}} let expressions, expressing a set of refactoring rules 
that eliminate several \ensuremath{\Conid{Haskell}} smells and solving tasks issued in the LDTA Tool Challenge~\cite{ldta} 
for name binding, type checking and desugaring of Oberon-0 programs. We 
benchmark our \ensuremath{\Conid{Haskell}} smell eliminator using 150 student Haskell projects totaling 82124 lines of code, 
from which we have eliminated 850 code smells.

\end{itemize}

This paper is organized as follows: Section~\ref{sec2} presents
generic zippers and describes in detail the zipper-based embedding
of strategic term rewriting. In Section~\ref{sec3}, we describe
zipper-based embedding of attribute grammars and we show how the two
techniques/embeddings can be easily combined. In Section~\ref{sec4} we
present our \ensuremath{\Conid{Ztrategic}} library and 
we use it in defining several usage
examples, such as refactorings of \ensuremath{\Conid{Haskell}} source code and name binding, 
type checking and desugaring Oberon-0 source code. Section~\ref{sec5} 
discusses related work, and in Section~\ref{sec6} we present our 
conclusions and future work.

\section{Zipper-Based Strategic Programming}
\label{sec2}

This section briefly describes functional Zippers~\cite{thezipper}
which are the building blocks of the embedding of strategic term
rewriting we introduce in this paper.  Before we present our
embedding in detail
later in the section,
 let us consider a motivating example we will use
throughout the paper. Consider the (sub)language of \ensuremath{\Conid{Let}} expressions
as incorporated in most functional languages, including
\ensuremath{\Conid{Haskell}}. Next, we show an example of a valid \ensuremath{\Conid{Haskell}} \ensuremath{\mathbf{let}}
expression 
and we define the heterogeneous data type \ensuremath{\Conid{Let}}, taken
from~\cite{scp2016}, that models such expressions in \ensuremath{\Conid{Haskell}} itself.

\begin{minipage}[t]{.4\textwidth}
\begin{hscode}\SaveRestoreHook
\column{B}{@{}>{\hspre}l<{\hspost}@{}}%
\column{6}{@{}>{\hspre}l<{\hspost}@{}}%
\column{12}{@{}>{\hspre}l<{\hspost}@{}}%
\column{E}{@{}>{\hspre}l<{\hspost}@{}}%
\>[B]{}\Varid{p}\mathrel{=}{}\<[6]%
\>[6]{}\mathbf{let}\;{}\<[12]%
\>[12]{}\Varid{a}\mathrel{=}\Varid{b}\mathbin{+}\mathrm{0}{}\<[E]%
\\
\>[12]{}\Varid{c}\mathrel{=}\mathrm{2}{}\<[E]%
\\
\>[12]{}\Varid{b}\mathrel{=}\mathbf{let}\;\Varid{c}\mathrel{=}\mathrm{3}\;\mathbf{in}\;\Varid{c}\mathbin{+}\Varid{c}{}\<[E]%
\\
\>[6]{}\mathbf{in}\;{}\<[12]%
\>[12]{}\Varid{a}\mathbin{+}\mathrm{7}\mathbin{-}\Varid{c}{}\<[E]%
\ColumnHook
\end{hscode}\resethooks
\end{minipage}
\begin{minipage}[t]{.4\textwidth}
\begin{hscode}\SaveRestoreHook
\column{B}{@{}>{\hspre}l<{\hspost}@{}}%
\column{7}{@{}>{\hspre}l<{\hspost}@{}}%
\column{13}{@{}>{\hspre}c<{\hspost}@{}}%
\column{13E}{@{}l@{}}%
\column{16}{@{}>{\hspre}l<{\hspost}@{}}%
\column{23}{@{}>{\hspre}l<{\hspost}@{}}%
\column{27}{@{}>{\hspre}l<{\hspost}@{}}%
\column{E}{@{}>{\hspre}l<{\hspost}@{}}%
\>[B]{}\mathbf{data}\;{}\<[7]%
\>[7]{}\Conid{Let}{}\<[13]%
\>[13]{}\mathrel{=}{}\<[13E]%
\>[16]{}\Conid{Let}\;\Conid{List}\;\Conid{Exp}{}\<[E]%
\\
\>[B]{}\mathbf{data}\;{}\<[7]%
\>[7]{}\Conid{List}{}\<[13]%
\>[13]{}\mathrel{=}{}\<[13E]%
\>[16]{}\Conid{NestedLet}\;{}\<[27]%
\>[27]{}\Conid{String}\;\Conid{Let}\;\Conid{List}{}\<[E]%
\\
\>[13]{}\mid {}\<[13E]%
\>[16]{}\Conid{Assign}\;{}\<[27]%
\>[27]{}\Conid{String}\;\Conid{Exp}\;\Conid{List}{}\<[E]%
\\
\>[13]{}\mid {}\<[13E]%
\>[16]{}\Conid{EmptyList}{}\<[E]%
\\
\>[B]{}\mathbf{data}\;{}\<[7]%
\>[7]{}\Conid{Exp}{}\<[13]%
\>[13]{}\mathrel{=}{}\<[13E]%
\>[16]{}\Conid{Add}\;{}\<[23]%
\>[23]{}\Conid{Exp}\;\Conid{Exp}{}\<[E]%
\\
\>[13]{}\mid {}\<[13E]%
\>[16]{}\Conid{Sub}\;{}\<[23]%
\>[23]{}\Conid{Exp}\;\Conid{Exp}{}\<[E]%
\\
\>[13]{}\mid {}\<[13E]%
\>[16]{}\Conid{Neg}\;{}\<[23]%
\>[23]{}\Conid{Exp}{}\<[E]%
\\
\>[13]{}\mid {}\<[13E]%
\>[16]{}\Conid{Var}\;{}\<[23]%
\>[23]{}\Conid{String}{}\<[E]%
\\
\>[13]{}\mid {}\<[13E]%
\>[16]{}\Conid{Const}\;{}\<[23]%
\>[23]{}\Conid{Int}{}\<[E]%
\ColumnHook
\end{hscode}\resethooks
\end{minipage}

Having introduced these data types, we can write \ensuremath{\Varid{p}} as a \ensuremath{\Conid{Haskell}}
(syntax) tree with type \ensuremath{\Conid{Let}}

\begin{hscode}\SaveRestoreHook
\column{B}{@{}>{\hspre}l<{\hspost}@{}}%
\column{6}{@{}>{\hspre}l<{\hspost}@{}}%
\column{11}{@{}>{\hspre}l<{\hspost}@{}}%
\column{23}{@{}>{\hspre}l<{\hspost}@{}}%
\column{28}{@{}>{\hspre}l<{\hspost}@{}}%
\column{34}{@{}>{\hspre}l<{\hspost}@{}}%
\column{E}{@{}>{\hspre}l<{\hspost}@{}}%
\>[B]{}\Varid{p}\mathbin{::}\Conid{Let}{}\<[E]%
\\
\>[B]{}\Varid{p}\mathrel{=}{}\<[6]%
\>[6]{}\Conid{Let}\;{}\<[11]%
\>[11]{}(\Conid{Assign}\;{}\<[23]%
\>[23]{}\text{\ttfamily \char34 a\char34}\;{}\<[28]%
\>[28]{}(\Conid{Add}\;(\Conid{Var}\;\text{\ttfamily \char34 b\char34})\;(\Conid{Const}\;\mathrm{0})){}\<[E]%
\\
\>[11]{}(\Conid{Assign}\;{}\<[23]%
\>[23]{}\text{\ttfamily \char34 c\char34}\;{}\<[28]%
\>[28]{}(\Conid{Const}\;\mathrm{2}){}\<[E]%
\\
\>[11]{}(\Conid{NestedLet}\;{}\<[23]%
\>[23]{}\text{\ttfamily \char34 b\char34}\;{}\<[28]%
\>[28]{}(\Conid{Let}\;{}\<[34]%
\>[34]{}(\Conid{Assign}\;\text{\ttfamily \char34 c\char34}\;(\Conid{Const}\;\mathrm{3}){}\<[E]%
\\
\>[34]{}\Conid{EmptyList})\;{}\<[E]%
\\
\>[34]{}(\Conid{Add}\;(\Conid{Var}\;\text{\ttfamily \char34 c\char34})\;(\Conid{Var}\;\text{\ttfamily \char34 c\char34}))){}\<[E]%
\\
\>[11]{}\Conid{EmptyList})))\;{}\<[E]%
\\
\>[11]{}(\Conid{Sub}\;(\Conid{Add}\;(\Conid{Var}\;\text{\ttfamily \char34 a\char34})\;(\Conid{Const}\;\mathrm{7}))\;(\Conid{Var}\;\text{\ttfamily \char34 c\char34})){}\<[E]%
\ColumnHook
\end{hscode}\resethooks

Consider now that we wish to implement a simple arithmetic optimizer
for our language. Let us start with a trivial optimization: the
elimination of additions with $0$. In this context, strategic term
rewriting is an extremely suitable formalism, since it provides a
solution that just defines the work to be done in the constructors
(tree nodes) of interest, and ``ignore" all the others. In our example,
the optimization is defined in \ensuremath{\Conid{Add}} nodes, and thus we express
the worker function as follows:

\begin{hscode}\SaveRestoreHook
\column{B}{@{}>{\hspre}l<{\hspost}@{}}%
\column{25}{@{}>{\hspre}l<{\hspost}@{}}%
\column{E}{@{}>{\hspre}l<{\hspost}@{}}%
\>[B]{}\Varid{expr}\mathbin{::}\Conid{Exp}\to \Conid{Maybe}\;\Conid{Exp}{}\<[E]%
\\
\>[B]{}\Varid{expr}\;(\Conid{Add}\;\Varid{e}\;(\Conid{Const}\;\mathrm{0})){}\<[25]%
\>[25]{}\mathrel{=}\Conid{Just}\;\Varid{e}{}\<[E]%
\\
\>[B]{}\Varid{expr}\;(\Conid{Add}\;(\Conid{Const}\;\mathrm{0})\;\Varid{e}){}\<[25]%
\>[25]{}\mathrel{=}\Conid{Just}\;\Varid{e}{}\<[E]%
\\
\>[B]{}\Varid{expr}\;\Varid{e}{}\<[25]%
\>[25]{}\mathrel{=}\Conid{Just}\;\Varid{e}{}\<[E]%
\ColumnHook
\end{hscode}\resethooks

The first two alternatives define the optimization (when either of the
sub-expressions of an Add expression is the constant $0$, then it
returns the other sub-expression), and the last one the default
behaviour for all other cases: it returns the original
expression. Because we also need to express transformations that may
fail (that is, do nothing), 
a type-specific transformation function returns a \ensuremath{\Conid{Maybe}}
result. 

This function applies to \ensuremath{\Conid{Exp}} nodes only. To express our \ensuremath{\Conid{Let}}
optimization, however, we need a generic mechanism that traverses a
\ensuremath{\Conid{Let}} program/tree, applying this function when visiting \ensuremath{\Conid{Add}}
expressions.  This is where strategic term rewriting comes to the
rescue: It provides recursion patterns (\textit{i.e.}, strategies) to
traverse the (generic) tree, like, for example, top-down or bottom-up
traversals. It also includes functions to apply a node specific
rewriting function (like \ensuremath{\Varid{expr}}) according to a given strategy. Next,
we show the strategic solution of our optimization where \ensuremath{\Varid{expr}} is
applied to the input tree in a full top-down strategy. This is a Type
Preserving (\ensuremath{\Conid{TP}}) transformation since the input and result trees have
the same type.

\begin{hscode}\SaveRestoreHook
\column{B}{@{}>{\hspre}l<{\hspost}@{}}%
\column{6}{@{}>{\hspre}l<{\hspost}@{}}%
\column{E}{@{}>{\hspre}l<{\hspost}@{}}%
\>[B]{}\Varid{opt}\mathbin{::}\textbf{Zipper}\;\Conid{Let}\to \Conid{Maybe}\;(\textbf{Zipper}\;\Conid{Let}){}\<[E]%
\\
\>[B]{}\Varid{opt}\;{}\<[6]%
\>[6]{}\Varid{t}\mathrel{=}\Varid{applyTP}\;(\mathit{full\_tdTP}\;\Varid{step})\;\Varid{t}{}\<[E]%
\\
\>[6]{}\mathbf{where}\;\Varid{step}\mathrel{=}\Varid{idTP}\mathbin{`\Varid{adhocTP}`}\Varid{expr}{}\<[E]%
\ColumnHook
\end{hscode}\resethooks

In fact, we have just presented our first zipper-based strategic
function.  Here, \ensuremath{\Varid{step}} is a transformation to be applied by function
\ensuremath{\Varid{applyTP}} to all nodes of the input tree \ensuremath{\Varid{t}} (of type \ensuremath{\textbf{Zipper}\;\Conid{Let}})
using a full top-down traversal scheme (function \ensuremath{\mathit{full\_tdTP}}). The
rewrite step behaves like the identity function (\ensuremath{\Varid{idTP}}) by default
with our \ensuremath{\Varid{expr}} function to perform the type-specific transformation,
and the \ensuremath{\Varid{adhocTP}} combinator joins them into a single function.

This strategic solution relies on our \ensuremath{\Conid{Ztrategic}}\footnote{The library and complete examples showed in this paper are available at 
\url{https://bitbucket.org/zenunomacedo/ztrategic/}}
library: a purely
functional embedding of strategic term rewriting in \ensuremath{\Conid{Haskell}}. In this
solution we clearly see that the traversal function \ensuremath{\mathit{full\_tdTP}} needs
to navigate on heterogeneous trees, as it is the case of the \ensuremath{\Conid{Let}}
expression \ensuremath{\Varid{p}}.  In a functional programming setting,
zippers~\cite{thezipper} provide a simple, but generic tree-walk
mechanism that we will use to embed strategic programming in
\ensuremath{\Conid{Haskell}}. In fact, our strategic combinators work with zippers as in
the definition of \ensuremath{\Varid{opt}}.  In the remaining of this section, we start
by briefly describing zippers, and, next, we present in detail the
embedding of strategies using this powerful mechanism.

\subsection{The Zipper Data Structure}

Zippers were introduced by Huet~\cite{thezipper} to represent a tree
together with a subtree that is the \emph{focus} of attention. During a
computation the focus may move left, up, down or right within the
tree.  Generic manipulation of a zipper is provided through a set of
predefined functions that allow access to all of the nodes of a tree
for inspection or modification.

A generic implementation of this concept is available as the \emph{generic
zipper} \ensuremath{\Conid{Haskell}} library~\cite{adams2010zippers}, which works for both
homogeneous and heterogeneous data types.  In order to illustrate the
use of zippers and its \ensuremath{\Conid{Haskell}} library, let us consider again the
tree used as an example for our \ensuremath{\Conid{Let}} program. 
This 
tree contains nodes of the types \ensuremath{\Conid{Let}}, \ensuremath{\Conid{List}} and \ensuremath{\Conid{Exp}}, and thus it
is an heterogeneous tree. Traditionally, a functional implementation
of a traversal of this tree would need three functions, one for each
different type that needs to be processed. Generic zippers, however,
provide a way to navigate in such heterogeneous data structures,
independently from the type of node it is traversing.

We build a zipper \ensuremath{\mathit{t_1}} from the previous \ensuremath{\Conid{Let}} expression \ensuremath{\Varid{p}} through
the use of the \ensuremath{\textbf{toZipper}\mathbin{::}\Conid{Data}\;\Varid{a}\Rightarrow \Varid{a}\to \textbf{Zipper}\;\Varid{a}} function. This
function produces a zipper out of any data type, requiring only that
the data types have an instance of the \ensuremath{\Conid{Data}} and \ensuremath{\Conid{Typeable}} type
classes\footnote{That can be easily obtained via the \ensuremath{\Conid{Haskell}} data
type \ensuremath{\mathbf{deriving}} mechanism.}.

\begin{hscode}\SaveRestoreHook
\column{B}{@{}>{\hspre}l<{\hspost}@{}}%
\column{E}{@{}>{\hspre}l<{\hspost}@{}}%
\>[B]{}\mathit{t_1}\mathrel{=}\textbf{toZipper}\;\Varid{p}{}\<[E]%
\ColumnHook
\end{hscode}\resethooks

We can navigate \ensuremath{\mathit{t_1}} using pre-defined functions from the zipper
library. The function \ensuremath{\textbf{down\textquoteright}} moves the focus down to the
leftmost child of a node, while \ensuremath{\textbf{down}} moves the focus to the
rightmost child instead.  Similarly, functions \ensuremath{\textbf{right}}, \ensuremath{\textbf{left}} and
\ensuremath{\textbf{up}}, move towards the corresponding directions. They have types:

\begin{hscode}\SaveRestoreHook
\column{B}{@{}>{\hspre}l<{\hspost}@{}}%
\column{30}{@{}>{\hspre}l<{\hspost}@{}}%
\column{E}{@{}>{\hspre}l<{\hspost}@{}}%
\>[B]{}\textbf{down},\textbf{down\textquoteright},\textbf{right},\textbf{left},\textbf{up}\mathbin{::}{}\<[30]%
\>[30]{}\textbf{Zipper}\;\Varid{a}\to \Conid{Maybe}\;(\textbf{Zipper}\;\Varid{a}){}\<[E]%
\ColumnHook
\end{hscode}\resethooks

Finally, the zipper function \ensuremath{\textbf{getHole}\mathbin{::}\Conid{Typeable}\;\Varid{b}\to \textbf{Zipper}\;\Varid{a}\to \Conid{Maybe}\;\Varid{b}} extracts the actual node the zipper is focusing on.  Using
these functions, we can freely navigate through this newly created
zipper. For example, consider our expression \ensuremath{\Varid{p}}. We can move the focus of the zipper towards the
\ensuremath{\Varid{b}\mathbin{+}\mathrm{0}} sub-expression as follows:

\begin{hscode}\SaveRestoreHook
\column{B}{@{}>{\hspre}l<{\hspost}@{}}%
\column{22}{@{}>{\hspre}c<{\hspost}@{}}%
\column{22E}{@{}l@{}}%
\column{25}{@{}>{\hspre}l<{\hspost}@{}}%
\column{E}{@{}>{\hspre}l<{\hspost}@{}}%
\>[B]{}\Varid{sumBZero}\mathbin{::}\Conid{Maybe}\;\Conid{Exp}{}\<[E]%
\\
\>[B]{}\Varid{sumBZero}\mathrel{=}(\textbf{getHole}{}\<[22]%
\>[22]{}\;.\;{}\<[22E]%
\>[25]{}\Varid{fromJust}\;.\;\textbf{right}{}\<[E]%
\\
\>[22]{}\;.\;{}\<[22E]%
\>[25]{}\Varid{fromJust}\;.\;\textbf{down\textquoteright}{}\<[E]%
\\
\>[22]{}\;.\;{}\<[22E]%
\>[25]{}\Varid{fromJust}\;.\;\textbf{down\textquoteright})\;\mathit{t_1}{}\<[E]%
\ColumnHook
\end{hscode}\resethooks

Because the navigation functions can fail, the data type \ensuremath{\Conid{Maybe}} is
used to make them total. To simplify our example we are unwrapping it
using the library function \ensuremath{\Varid{fromJust}}.

To avoid the repeated use of \ensuremath{\Varid{fromJust}} and to define total
functions, which also express a more natural top-down
writing/reading of the navigation on trees, we can rewrite these
functions using the monadic do-notation\footnote{A pure monadic
definition can also be used, which make these definitions even
simpler. In section~\ref{sec5} we will show how to obtain them.}, as
follows:

\begin{hscode}\SaveRestoreHook
\column{B}{@{}>{\hspre}l<{\hspost}@{}}%
\column{14}{@{}>{\hspre}l<{\hspost}@{}}%
\column{18}{@{}>{\hspre}l<{\hspost}@{}}%
\column{32}{@{}>{\hspre}l<{\hspost}@{}}%
\column{E}{@{}>{\hspre}l<{\hspost}@{}}%
\>[B]{}\Varid{sumBZero'}\mathbin{::}\Conid{Maybe}\;\Conid{Exp}{}\<[E]%
\\
\>[B]{}\Varid{sumBZero'}\mathrel{=}{}\<[14]%
\>[14]{}\mathbf{do}\;{}\<[18]%
\>[18]{}\mathit{t_2}\leftarrow \textbf{down\textquoteright}\;{}\<[32]%
\>[32]{}\mathit{t_1}{}\<[E]%
\\
\>[18]{}\mathit{t_3}\leftarrow \textbf{down\textquoteright}\;{}\<[32]%
\>[32]{}\mathit{t_2}{}\<[E]%
\\
\>[18]{}\mathit{t_4}\leftarrow \textbf{right}\;{}\<[32]%
\>[32]{}\mathit{t_3}{}\<[E]%
\\
\>[18]{}\textbf{getHole}\;\mathit{t_4}{}\<[E]%
\ColumnHook
\end{hscode}\resethooks

The zipper library also contains functions for the transformation of the
data structure being traversed. The function \ensuremath{\textbf{trans}\mathbin{::}\Conid{GenericT}\to \textbf{Zipper}\;\Varid{a}\to \textbf{Zipper}\;\Varid{a}} applies a generic transformation to the node the
zipper is currently pointing towards, while  \ensuremath{\textbf{transM}\mathbin{::}\Conid{GenericM}\;\Varid{m}\to \textbf{Zipper}\;\Varid{a}\to \Varid{m}\;(\textbf{Zipper}\;\Varid{a})} applies a generic monadic
transformation.


In order to show a zipper-based transformation, let us consider 
that we wish to increment a constant in a \ensuremath{\Conid{Let}} expression.
We begin by defining a trivial function that increments constants,

\begin{hscode}\SaveRestoreHook
\column{B}{@{}>{\hspre}l<{\hspost}@{}}%
\column{E}{@{}>{\hspre}l<{\hspost}@{}}%
\>[B]{}\Varid{incConstant}\mathbin{::}\Conid{Exp}\to \Conid{Exp}{}\<[E]%
\\
\>[B]{}\Varid{incConstant}\;(\Conid{Const}\;\Varid{n})\mathrel{=}\Conid{Const}\;(\Varid{n}\mathbin{+}\mathrm{1}){}\<[E]%
\ColumnHook
\end{hscode}\resethooks

Since \ensuremath{\Varid{incConstant}} works on type \ensuremath{\Conid{Exp}} only, we use the generic
function \ensuremath{\Varid{mkT}} (from the generics library~\cite{syb}) to generalize
this type-specific function to all types.

\begin{hscode}\SaveRestoreHook
\column{B}{@{}>{\hspre}l<{\hspost}@{}}%
\column{E}{@{}>{\hspre}l<{\hspost}@{}}%
\>[B]{}\Varid{incConstantG}\mathbin{::}\Conid{GenericT}{}\<[E]%
\\
\>[B]{}\Varid{incConstantG}\mathrel{=}\Varid{mkT}\;\Varid{incConstant}{}\<[E]%
\ColumnHook
\end{hscode}\resethooks

This function has type \ensuremath{\Conid{GenericT}} (meaning \emph{Generic Transformation})
that is the required type of \ensuremath{\textbf{trans}}. To transform the assignment \ensuremath{\Varid{c}\mathrel{=}\mathrm{2}}
(in \ensuremath{\Varid{p}}) to \ensuremath{\Varid{c}\mathrel{=}\mathrm{3}} we just have to navigate to the desired constant
and then apply \ensuremath{\textbf{trans}}, as follows:

\begin{hscode}\SaveRestoreHook
\column{B}{@{}>{\hspre}l<{\hspost}@{}}%
\column{13}{@{}>{\hspre}l<{\hspost}@{}}%
\column{27}{@{}>{\hspre}l<{\hspost}@{}}%
\column{E}{@{}>{\hspre}l<{\hspost}@{}}%
\>[B]{}\Varid{incrC}\mathbin{::}\Conid{Maybe}\;(\textbf{Zipper}\;\Conid{Let}){}\<[E]%
\\
\>[B]{}\Varid{incrC}\mathrel{=}\mathbf{do}\;{}\<[13]%
\>[13]{}\mathit{t_2}\leftarrow \textbf{down\textquoteright}\;{}\<[27]%
\>[27]{}\mathit{t_1}{}\<[E]%
\\
\>[13]{}\mathit{t_3}\leftarrow \textbf{down\textquoteright}\;{}\<[27]%
\>[27]{}\mathit{t_2}{}\<[E]%
\\
\>[13]{}\mathit{t_4}\leftarrow \textbf{right}\;{}\<[27]%
\>[27]{}\mathit{t_3}{}\<[E]%
\\
\>[13]{}\mathit{t_5}\leftarrow \textbf{right}\;{}\<[27]%
\>[27]{}\mathit{t_4}{}\<[E]%
\\
\>[13]{}\mathit{t_6}\leftarrow \textbf{down\textquoteright}\;{}\<[27]%
\>[27]{}\mathit{t_5}{}\<[E]%
\\
\>[13]{}\mathit{t_7}\leftarrow \textbf{right}\;{}\<[27]%
\>[27]{}\mathit{t_6}{}\<[E]%
\\
\>[13]{}\Varid{return}\;(\textbf{trans}\;\Varid{incConstantG}\;\mathit{t_7}){}\<[E]%
\ColumnHook
\end{hscode}\resethooks

In fact, generic zippers is a simple, but very expressive technique to
navigate in heterogeneous data structures. Since strategic term
rewriting relies on generic traversals of trees, and on the
transformation of specific nodes, zippers provide the necessary
machinery to embed strategic programming in \ensuremath{\Conid{Haskell}}, as we will show
in the next section. It should also me mentioned that zippers also
provide a powerful embedding of attribute grammars in
\ensuremath{\Conid{Haskell}}~\cite{zipperAG,scp2016,memoAG19}. In Section~\ref{sec3}
we will show how these two language engineering techniques/embeddings
can be easily combined as a result of being expressed on the same
setting; \textit{i.e.} via zippers.

\subsection{Strategic Programming}

In this section we introduce the embedding of strategic programming
using generic zippers. Our embedding directly follows the work of
Laemmel and Visser~\cite{strafunski} on the Strafunski
library~\cite{StrafunskiAppLetter}. Before we present the powerful and
reusable strategic functions providing control on tree traversals,
such as top-down, bottom-up, innermost, etc., let us show some simple
basic combinators that work at the zipper level, and are the building
blocks of our embedding.

We start by defining a function that expresses how a given
transformation function is elevated to the zipper level. In other
words, we define how a function that is supposed to operate directly
on one data type is converted into a transformation that operates on a
zipper.

\begin{hscode}\SaveRestoreHook
\column{B}{@{}>{\hspre}l<{\hspost}@{}}%
\column{13}{@{}>{\hspre}l<{\hspost}@{}}%
\column{E}{@{}>{\hspre}l<{\hspost}@{}}%
\>[B]{}\Varid{zTryApplyM}{}\<[13]%
\>[13]{}\mathbin{::}(\Conid{Typeable}\;\Varid{a},\Conid{Typeable}\;\Varid{b})\Rightarrow (\Varid{a}\to \Conid{Maybe}\;\Varid{b})\to \Conid{TP}\;\Varid{c}{}\<[E]%
\\
\>[B]{}\Varid{zTryApplyM}\;\Varid{f}\mathrel{=}\textbf{transM}\;(\Varid{join}\;.\;\Varid{cast}\;.\;\Varid{f}\;.\;\Varid{fromJust}\;.\;\Varid{cast}){}\<[E]%
\ColumnHook
\end{hscode}\resethooks

The definition of \ensuremath{\Varid{zTryApplyM}} relies on transformation on zippers,
thus reusing the generic zipper library \ensuremath{\textbf{transM}} function.
To build a valid transformation for the \ensuremath{\textbf{transM}} function, we
use the \ensuremath{\Varid{cast}\mathbin{::}\Varid{a}\to \Conid{Maybe}\;\Varid{b}} function, that tries to cast a given
data from type \ensuremath{\Varid{a}} to type \ensuremath{\Varid{b}}. In this case, we use it to cast the
data the zipper is focused on into a type our original transformation
\ensuremath{\Varid{f}} can be applied to.  Then, function \ensuremath{\Varid{f}} is applied and its result
is cast back to its original type. Should any of these casts, or the
function \ensuremath{\Varid{f}} itself, fail, the failure propagates and the resulting
zipper transformation will also fail. The use of the monadic version
of the zipper generic transformation guarantees the handling of such
partiality.  It should be noticed that failure can occur in two
situations: the type cast fails when the types do not match. Moreover,
the function \ensuremath{\Varid{f}} fails when the function itself dictates that no
change is to be applied.  Signaling failure in the application of
transformations is important for strategies where a transformation is
applied once, only.


\ensuremath{\Varid{zTryApplyM}} returns a \ensuremath{\Conid{TP}\;\Varid{c}}, in which \ensuremath{\Conid{TP}} is a type for specifying
Type-Preserving transformations on zippers, and \ensuremath{\Varid{c}} is the type of the
zipper. For example, if we are applying transformations on a zipper
built upon the \ensuremath{\Conid{Let}} data type, then those transformations are of type
\ensuremath{\Conid{TP}\;\Conid{Let}}. Very much like Strafunski, we also introduce the type \ensuremath{\Conid{TU}\;\Varid{m}\;\Varid{d}} for Type-Unifying operations, which aim to gather data of type \ensuremath{\Varid{d}}
into the data structure \ensuremath{\Varid{m}}. For example, to collect in a list all the
defined names in a \ensuremath{\Conid{Let}} expression, the corresponding type-unifying
strategy would be of type \ensuremath{\Conid{TU}\;[\mskip1.5mu \mskip1.5mu]\;\Conid{String}}. We will present such a
transformation and implement it later in this section.

Next, we define a combinator to compose two transformations, building
a more complex zipper transformation that tries to apply each of the
initial transformations in sequence. Because each of the
transformations may fail, we have to skip transformations that fail.

\begin{hscode}\SaveRestoreHook
\column{B}{@{}>{\hspre}l<{\hspost}@{}}%
\column{10}{@{}>{\hspre}c<{\hspost}@{}}%
\column{10E}{@{}l@{}}%
\column{14}{@{}>{\hspre}l<{\hspost}@{}}%
\column{43}{@{}>{\hspre}l<{\hspost}@{}}%
\column{E}{@{}>{\hspre}l<{\hspost}@{}}%
\>[B]{}\Varid{adhocTP}{}\<[10]%
\>[10]{}\mathbin{::}{}\<[10E]%
\>[14]{}(\Conid{Typeable}\;\Varid{a},\Conid{Typeable}\;\Varid{b})\Rightarrow {}\<[43]%
\>[43]{}\Conid{TP}\;\Varid{e}\to (\Varid{a}\to \Conid{Maybe}\;\Varid{b})\to \Conid{TP}\;\Varid{e}{}\<[E]%
\\
\>[B]{}\Varid{adhocTP}\;\Varid{f}\;\Varid{g}\;\Varid{z}\mathrel{=}\Varid{maybeKeep}\;\Varid{f}\;(\Varid{zTryApplyM}\;\Varid{g})\;\Varid{z}{}\<[E]%
\ColumnHook
\end{hscode}\resethooks

The \ensuremath{\Varid{adhocTP}} function receives transformations \ensuremath{\Varid{f}} and \ensuremath{\Varid{g}} as
parameters, as well as zipper \ensuremath{\Varid{z}}. It converts \ensuremath{\Varid{g}} into a zipper
transformation, and it uses the auxiliary function \ensuremath{\Varid{maybeKeep}} to try
to apply each of the transformations to the zipper \ensuremath{\Varid{z}}, ignoring the
transformations that fail.  Note that \ensuremath{\Varid{f}} is of type \ensuremath{\Conid{TP}\;\Varid{e}}, meaning
it is a transformation on zippers, while \ensuremath{\Varid{g}} is a normal \ensuremath{\Conid{Haskell}}
function. Because \ensuremath{\Varid{g}} is a non-zipper based function, \ensuremath{\Varid{adhocTP}} allows
the definition of transformations where we use simple (\textit{i.e.}
non-zipper) \ensuremath{\Conid{Haskell}} functions. Next, we show an example of the use of \ensuremath{\Varid{adhocTP}},
written as an infix operator, which combines the zipper function
\ensuremath{\Varid{failTP}} with our basic transformation \ensuremath{\Varid{expr}} function:

\begin{hscode}\SaveRestoreHook
\column{B}{@{}>{\hspre}l<{\hspost}@{}}%
\column{E}{@{}>{\hspre}l<{\hspost}@{}}%
\>[B]{}\Varid{step}\mathrel{=}\Varid{failTP}\mathbin{`\Varid{adhocTP}`}\Varid{expr}{}\<[E]%
\ColumnHook
\end{hscode}\resethooks

Thus, we do not need to express type-specific transformations as
functions that work on zippers. It is the use of \ensuremath{\Varid{zTryApplyM}} in
\ensuremath{\Varid{adhocTP}} that transforms a normal \ensuremath{\Conid{Haskell}} function (\ensuremath{\Varid{expr}} in this
case) to a zipper one, hidden from these definitions.

The function \ensuremath{\Varid{failTP}} is a pre-defined transformation that always
fails and \ensuremath{\Varid{idTP}} is the identity transformation that always
succeeds. They provide the basis for construction of complex
transformations through composition.  We omit here their simple
definitions.

The functions we have presented already allow the definition of
arbitrarily complex transformations for zippers. Such transformations,
however, are always applied on the node the zipper is focusing on. Let
us consider a combinator that does navigate in the zipper.

\begin{hscode}\SaveRestoreHook
\column{B}{@{}>{\hspre}l<{\hspost}@{}}%
\column{19}{@{}>{\hspre}l<{\hspost}@{}}%
\column{28}{@{}>{\hspre}l<{\hspost}@{}}%
\column{E}{@{}>{\hspre}l<{\hspost}@{}}%
\>[B]{}\Varid{allTPright}\mathbin{::}\Conid{TP}\;\Varid{a}\to \Conid{TP}\;\Varid{a}{}\<[E]%
\\
\>[B]{}\Varid{allTPright}\;\Varid{f}\;\Varid{z}\mathrel{=}{}\<[19]%
\>[19]{}\mathbf{case}\;\textbf{right}\;\Varid{z}\;\mathbf{of}{}\<[E]%
\\
\>[19]{}\Conid{Nothing}{}\<[28]%
\>[28]{}\to \Varid{return}\;\Varid{z}{}\<[E]%
\\
\>[19]{}\Conid{Just}\;\Varid{r}{}\<[28]%
\>[28]{}\to \Varid{fmap}\;(\Varid{fromJust}\;.\;\textbf{left})\;(\Varid{f}\;\Varid{r}){}\<[E]%
\ColumnHook
\end{hscode}\resethooks

The code presented above is a combinator that, given a type-preserving
transformation \ensuremath{\Varid{f}} for zipper \ensuremath{\Varid{z}}, will attempt to apply \ensuremath{\Varid{f}} to the
node that is located to the right of the node the zipper is pointing
towards. To do this, the zipper function \ensuremath{\textbf{right}} is used to try to
navigate to the right; if it fails, we return the original zipper. If
it succeeds, we apply transformation \ensuremath{\Varid{f}} and then we navigate \ensuremath{\textbf{left}}
again. There is a similar combinator named \ensuremath{\Varid{allTPdown}} that will
perform the same logic but by navigating downwards and then upwards.

With all these tools at our disposal, we can define generic traversal schemes by combining them. Next, we define the traversal scheme used in the function \ensuremath{\Varid{opt}} we defined at the start of the section. This traversal scheme navigates through the whole data structure, in a top-down approach. 

\begin{hscode}\SaveRestoreHook
\column{B}{@{}>{\hspre}l<{\hspost}@{}}%
\column{16}{@{}>{\hspre}l<{\hspost}@{}}%
\column{28}{@{}>{\hspre}l<{\hspost}@{}}%
\column{62}{@{}>{\hspre}l<{\hspost}@{}}%
\column{E}{@{}>{\hspre}l<{\hspost}@{}}%
\>[B]{}\mathit{full\_tdTP}\mathbin{::}\Conid{TP}\;\Varid{a}\to \Conid{TP}\;\Varid{a}{}\<[E]%
\\
\>[B]{}\mathit{full\_tdTP}\;\Varid{f}\mathrel{=}{}\<[16]%
\>[16]{}\Varid{allTPdown}\;{}\<[28]%
\>[28]{}(\mathit{full\_tdTP}\;\Varid{f})\mathbin{`\Varid{seqTP}`}\Varid{allTPright}\;{}\<[62]%
\>[62]{}(\mathit{full\_tdTP}\;\Varid{f})\mathbin{`\Varid{seqTP}`}\Varid{f}{}\<[E]%
\ColumnHook
\end{hscode}\resethooks

We skip the explanation of the \ensuremath{\Varid{seqTP}} operator as it is relatively
similar to the \ensuremath{\Varid{adhocTP}} operator we described before, albeit simpler;
we interpret this as a sequence operator. This function receives as
input a type-preserving transformation \ensuremath{\Varid{f}}, and it (reading the code
from right to left) applies it to the focused node itself, then to the
nodes below the currently focused node, then to the nodes to the right
of the focused node. To apply this transformation to the nodes below
the current node, for example, we use the \ensuremath{\Varid{allTPdown}} combinator we
defined before, and we recursively apply \ensuremath{\mathit{full\_tdTP}\;\Varid{f}} to the node
below. The same logic applies in regards to navigating to the right.

We can define several traversal schemes similar to this one by changing the combinators used, or their sequence. For example, by inverting the order in which the combinators are sequenced, we define a bottom-up traversal. By using different combinators, we can define choice, allowing for partial traversals in the data structure. 

In fact, previously we defined a rewrite strategy where we use
\ensuremath{\mathit{full\_tdTP}} to define a full, top-down traversal, which is not
ideal. Because we intend to optimize \ensuremath{\Conid{Exp}} nodes, changing one node
might make it possible to optimize the node above, which would have
already been processed in a top-down traversal. Instead, we define a
different traversal scheme, for repeated application of a
transformation until a fixed point is reached:

\begin{hscode}\SaveRestoreHook
\column{B}{@{}>{\hspre}l<{\hspost}@{}}%
\column{12}{@{}>{\hspre}l<{\hspost}@{}}%
\column{14}{@{}>{\hspre}c<{\hspost}@{}}%
\column{14E}{@{}l@{}}%
\column{17}{@{}>{\hspre}l<{\hspost}@{}}%
\column{E}{@{}>{\hspre}l<{\hspost}@{}}%
\>[B]{}\Varid{innermost}{}\<[12]%
\>[12]{}\mathbin{::}\Conid{TP}\;\Varid{a}\to \Conid{TP}\;\Varid{a}{}\<[E]%
\\
\>[B]{}\Varid{innermost}\;\Varid{s}{}\<[14]%
\>[14]{}\mathrel{=}{}\<[14E]%
\>[17]{}\Varid{repeatTP}\;(\mathit{once\_buTP}\;\Varid{s}){}\<[E]%
\ColumnHook
\end{hscode}\resethooks

We omit the definitions of \ensuremath{\mathit{once\_buTP}} and \ensuremath{\Varid{repeatTP}} as they are
similar to the definitions presented already. The combinator \ensuremath{\Varid{repeatTP}}
applies a given transformation repeatedly until a fixed point is
reached, that is, until the data structure stops being changed by the
transformation. The transformation being applied repeatedly is defined
with the \ensuremath{\mathit{once\_buTP}} combinator, which applies \ensuremath{\Varid{s}} once, anywhere on
the data structure. When the application \ensuremath{\mathit{once\_buTP}} fails, \ensuremath{\Varid{repeatTP}}
understands a fixed point is reached. Because the \ensuremath{\mathit{once\_buTP}}
bottom-up combinator is used, the traversal scheme is \ensuremath{\Varid{innermost}}, since
it always prioritizes the innermost nodes. The \ensuremath{\Varid{outermost}} function
is defined in a similar way, but using the \ensuremath{\mathit{once\_tdTP}} combinator
instead.

Let us return to our \ensuremath{\Conid{Let}} running example. Obviously there are more
arithmetic rules that we may use to optimize let expressions. In
Figure~\ref{rules} we present the rules literally taken
from~\cite{strategicAG}.

\begin{figure}
\begin{align}\label{eq:rules}
add(e, const(0)) &\rightarrow e \\
add(const(0), e) &\rightarrow e \\
add(const(a), const(b)) &\rightarrow const(a+b) \\
sub(e1, e2) &\rightarrow add(e1, neg(e2)) \\
neg(neg(e)) &\rightarrow e \\
neg(const(a)) &\rightarrow const(-a) \\
var(id) \mid (id, just(e)) \in env &\rightarrow e
\end{align}
\caption{Optimization Rules}
\label{rules}
\end{figure}

In our previous definition of the function \ensuremath{\Varid{expr}}, we already defined
rewriting rules for optimizations $1$ and $2$. Rules $3$ through $6$
can also be trivially defined in \ensuremath{\Conid{Haskell}}:

\begin{hscode}\SaveRestoreHook
\column{B}{@{}>{\hspre}l<{\hspost}@{}}%
\column{33}{@{}>{\hspre}l<{\hspost}@{}}%
\column{E}{@{}>{\hspre}l<{\hspost}@{}}%
\>[B]{}\Varid{expr}\mathbin{::}\Conid{Exp}\to \Conid{Maybe}\;\Conid{Exp}{}\<[E]%
\\
\>[B]{}\Varid{expr}\;(\Conid{Add}\;\Varid{e}\;(\Conid{Const}\;\mathrm{0})){}\<[33]%
\>[33]{}\mathrel{=}\Conid{Just}\;\Varid{e}{}\<[E]%
\\
\>[B]{}\Varid{expr}\;(\Conid{Add}\;(\Conid{Const}\;\mathrm{0})\;\Varid{t}){}\<[33]%
\>[33]{}\mathrel{=}\Conid{Just}\;\Varid{t}{}\<[E]%
\\
\>[B]{}\Varid{expr}\;(\Conid{Add}\;(\Conid{Const}\;\Varid{a})\;(\Conid{Const}\;\Varid{b})){}\<[33]%
\>[33]{}\mathrel{=}\Conid{Just}\;(\Conid{Const}\;(\Varid{a}\mathbin{+}\Varid{b})){}\<[E]%
\\
\>[B]{}\Varid{expr}\;(\Conid{Sub}\;\Varid{a}\;\Varid{b}){}\<[33]%
\>[33]{}\mathrel{=}\Conid{Just}\;(\Conid{Add}\;\Varid{a}\;(\Conid{Neg}\;\Varid{b})){}\<[E]%
\\
\>[B]{}\Varid{expr}\;(\Conid{Neg}\;(\Conid{Neg}\;\Varid{f})){}\<[33]%
\>[33]{}\mathrel{=}\Conid{Just}\;\Varid{f}{}\<[E]%
\\
\>[B]{}\Varid{expr}\;(\Conid{Neg}\;(\Conid{Const}\;\Varid{n})){}\<[33]%
\>[33]{}\mathrel{=}\Conid{Just}\;(\Conid{Const}\;(\mathbin{-}\Varid{n})){}\<[E]%
\\
\>[B]{}\Varid{expr}\;\anonymous {}\<[33]%
\>[33]{}\mathrel{=}\Conid{Nothing}{}\<[E]%
\ColumnHook
\end{hscode}\resethooks


Rule $7$, however, is context dependent and it is not easily expressed
within strategic term rewriting.  In fact, this rule requires that
the environment, where a name is used, is computed first (according to
the scope rules of the \ensuremath{\Conid{Let}} language). We will return to this rule in
Section~\ref{sec3}.

Having expressed all rewriting rules in function \ensuremath{\Varid{expr}}, now we need
to use our strategic combinators that navigate in the tree while
applying the rules.  To guarantee that all the possible optimizations
are applied we use an \ensuremath{\Varid{innermost}} traversal scheme. Thus, our
optimization is expressed as:

\begin{hscode}\SaveRestoreHook
\column{B}{@{}>{\hspre}l<{\hspost}@{}}%
\column{7}{@{}>{\hspre}l<{\hspost}@{}}%
\column{12}{@{}>{\hspre}l<{\hspost}@{}}%
\column{E}{@{}>{\hspre}l<{\hspost}@{}}%
\>[B]{}\Varid{opt'}{}\<[7]%
\>[7]{}\mathbin{::}\textbf{Zipper}\;\Conid{Let}\to \Conid{Maybe}\;(\textbf{Zipper}\;\Conid{Let}){}\<[E]%
\\
\>[B]{}\Varid{opt'}\;{}\<[7]%
\>[7]{}\Varid{t}\mathrel{=}{}\<[12]%
\>[12]{}\Varid{applyTP}\;(\Varid{innermost}\;\Varid{step})\;\Varid{t}{}\<[E]%
\\
\>[7]{}\mathbf{where}\;\Varid{step}\mathrel{=}\Varid{failTP}\mathbin{`\Varid{adhocTP}`}\Varid{expr}{}\<[E]%
\ColumnHook
\end{hscode}\resethooks

Function \ensuremath{\Varid{opt'}} combines all the steps we have built until now. We
define an auxiliary function \ensuremath{\Varid{step}}, which is the composition of the
\ensuremath{\Varid{failTP}} default failing strategy with \ensuremath{\Varid{expr}}, the optimization
function; we compose them with \ensuremath{\Varid{adhocTP}}.  Our resulting
Type-Preserving strategy will be \ensuremath{\Varid{innermost}\;\Varid{step}}, which applies
\ensuremath{\Varid{step}} to the zipper as many times as possible until a fixed-point is
reached.  The use of \ensuremath{\Varid{failTP}} as the default strategy is required, as
\ensuremath{\Varid{innermost}} reaches the fixed-point when \ensuremath{\Varid{step}} fails.
If we use \ensuremath{\Varid{idTP}} instead, \ensuremath{\Varid{step}} always succeeds, resulting
in an infinite loop. We apply this strategy using the function 
\ensuremath{\Varid{applyTP}\mathbin{::}\Conid{TP}\;\Varid{c}\to \textbf{Zipper}\;\Varid{c}\to \Conid{Maybe}\;(\textbf{Zipper}\;\Varid{c})}, 
which effectively applies a strategy to a zipper. This function is defined
in our library, but we omit the code as it is trivial. 

Next, we show an example using a Type-Unifying strategy. More
concretely, we define a function \ensuremath{\Varid{names}} that collects all defined
names in a \ensuremath{\Conid{Let}} expression. First, we define a function \ensuremath{\Varid{select}} that focus on
the \ensuremath{\Conid{Let}} tree nodes where names are defined, namely, \ensuremath{\Conid{Assign}} and
\ensuremath{\Conid{NestedLet}}. This function returns a singleton list (with the defined
name) when applied to these nodes, and an empty list in the other
cases.

\begin{hscode}\SaveRestoreHook
\column{B}{@{}>{\hspre}l<{\hspost}@{}}%
\column{19}{@{}>{\hspre}l<{\hspost}@{}}%
\column{27}{@{}>{\hspre}l<{\hspost}@{}}%
\column{E}{@{}>{\hspre}l<{\hspost}@{}}%
\>[B]{}\Varid{select}\mathbin{::}\Conid{List}\to [\mskip1.5mu \Conid{String}\mskip1.5mu]{}\<[E]%
\\
\>[B]{}\Varid{select}\;(\Conid{Assign}\;{}\<[19]%
\>[19]{}\Varid{s}\;\anonymous \;\anonymous ){}\<[27]%
\>[27]{}\mathrel{=}[\mskip1.5mu \Varid{s}\mskip1.5mu]{}\<[E]%
\\
\>[B]{}\Varid{select}\;(\Conid{NestedLet}\;\Varid{s}\;\anonymous \;\anonymous ){}\<[27]%
\>[27]{}\mathrel{=}[\mskip1.5mu \Varid{s}\mskip1.5mu]{}\<[E]%
\\
\>[B]{}\Varid{select}\;\anonymous {}\<[27]%
\>[27]{}\mathrel{=}[\mskip1.5mu \mskip1.5mu]{}\<[E]%
\ColumnHook
\end{hscode}\resethooks

Now, \ensuremath{\Varid{names}} is a Type-Unifying function that traverses a given \ensuremath{\Conid{Let}}
tree (inside a zipper, in our case), and produces a list with the
declared names.

\begin{hscode}\SaveRestoreHook
\column{B}{@{}>{\hspre}l<{\hspost}@{}}%
\column{8}{@{}>{\hspre}l<{\hspost}@{}}%
\column{E}{@{}>{\hspre}l<{\hspost}@{}}%
\>[B]{}\Varid{names}{}\<[8]%
\>[8]{}\mathbin{::}\textbf{Zipper}\;\Conid{Let}\to [\mskip1.5mu \Conid{String}\mskip1.5mu]{}\<[E]%
\\
\>[B]{}\Varid{names}\;{}\<[8]%
\>[8]{}\Varid{r}\mathrel{=}\Varid{applyTU}\;(\mathit{full\_tdTU}\;\Varid{step})\;\Varid{r}{}\<[E]%
\\
\>[8]{}\mathbf{where}\;\Varid{step}\mathrel{=}\Varid{failTU}\mathbin{`\Varid{adhocTU}`}\Varid{select}{}\<[E]%
\ColumnHook
\end{hscode}\resethooks

The traversal strategy influences the order of the names in the
resulting list. We use a top-down traversal so that the list result
follows the order of the input. This is to say that \ensuremath{\Varid{names}\;\mathit{t_1}\equiv [\mskip1.5mu \text{\ttfamily \char34 a\char34},\text{\ttfamily \char34 c\char34},\text{\ttfamily \char34 b\char34},\text{\ttfamily \char34 c\char34}\mskip1.5mu]} (a bottom-up strategy produces the reverse of this
list).

As we have shown, our strategic term rewriting functions rely on
zippers built upon the data (trees) to be traversed. This results in
strategic functions that can easily be combined with a zipper-based
embedding of attribute grammars~\cite{scp2016,memoAG19}, since both
functions/embedding work on zippers. In the next section we present in
detail the zipping of strategies and AGs.

\section{Strategic Attribute Grammars}
\label{sec3}

Zipper-based strategic term rewriting provides a powerful mechanism
to express tree transformations. There are, however, transformations
that rely on context information that needs to be collected 
before the transformation can be applied. Our optimization rule $7$ of Figure~\ref{rules} is
such an example.

In this section we will briefly explain the Zipper-based embedding of attribute grammars, through the \ensuremath{\Conid{Let}} example.
Then, we are going to introduce how to combine strategies and AGs, ending with an implementation of rule 7.

\subsection{Zipper-based Attribute Grammars}

The attribute grammar formalism is particularly suitable to specify
language-based algorithms, where context information needs to be first
collected before it can be used. Language-based algorithms such as
name analysis~\cite{scp2016}, pretty printing~\cite{Swierstra1999Combinator},
type inference~\cite{Middelkoop2010TypeInference}, etc.  are elegantly
specified using attribute grammars.

Our running example is no exception and the name analysis task of
\ensuremath{\Conid{Let}} is a non-trivial one.  While being a concise example, it holds
central characteristics of software languages, such as (nested)
block-based structures and mandatory but unique declarations of
names. In addition, the semantics of \ensuremath{\Conid{Let}} does not force a
declare-before-use discipline, meaning that a variable can be declared
after its first use. Consequently, a conventional  implementation of name analysis
naturally leads to a processor that traverses each block twice: once
for processing the declarations of names and constructing an
environment and a second time to process the uses of names (using the
computed environment) in order to check for the use of non-declared
identifiers. The uniqueness of identifiers is efficiently checked in
the first traversal: for each newly encountered name it is checked
whether that it has already been declared at the same lexical level
(block). As a consequence, semantic errors resulting from duplicate
definitions are computed during the first traversal and errors
resulting from missing declarations, in the second one. In fact,
expressing this straightforward algorithm is a complex task in most
programming paradigms, since they require a complex scheduling of tree
traversals\footnote{Note that only after building the environment of an
outer block, nested ones can be traversed for the first time: they
inherited that environment. Thus, traversals are intermingled.}, and
intrusive code may be needed to pass information computed in
one traversal to a specific node and used in a following
one\footnote{This is the case when we wish to produce a list of errors
that follows the sequential structure of the input program~\cite{Saraiva99}.}.

In the attribute grammar paradigm, the programmer does not need to be
concerned with scheduling of traversals, nor the use of intrusive code
to glue traversals together. As a consequence, AG writers do not need
to transform/adapt the clear and elegant algorithms in order to avoid
those issues.

Attribute grammars are context-free~\cite{knuth1968semantics}. That is
to say that the root symbol does not have inherited attributes. Since
this not the case of symbol \ensuremath{\Conid{Let}} (due to its nested occurrence in the
grammar/data types), we will add a root symbol to our grammar as
follows:

\begin{hscode}\SaveRestoreHook
\column{B}{@{}>{\hspre}l<{\hspost}@{}}%
\column{E}{@{}>{\hspre}l<{\hspost}@{}}%
\>[B]{}\mathbf{data}\;\Conid{Root}\mathrel{=}\Conid{Root}\;\Conid{Let}{}\<[E]%
\ColumnHook
\end{hscode}\resethooks

\begin{figure*}[h]
\includegraphics[width=\textwidth,keepaspectratio]{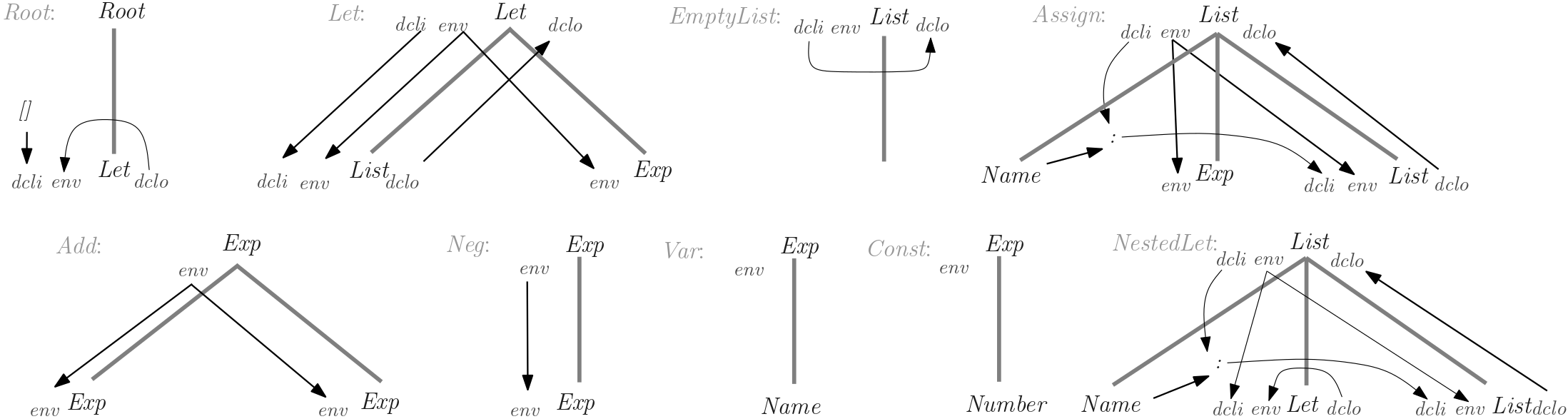}
\caption{Attribute Grammar Specifying the Scope Rules of \ensuremath{\Conid{Let}}}
\label{fig:AG}
\end{figure*}

Instead of presenting the formal AG definition of \ensuremath{\Conid{Let}}, we will adopt
a visual notation that is often used by AG writers to sketch a first
draft of their grammars.  The scope rules of \ensuremath{\Conid{Let}} are visually
expressed in Figure~\ref{fig:AG}. 

The diagrams in the figure are read as follows. For each production
(labeled by its name) we have the type of the production above and
below those of its children. To the left of each symbol we have the
so-called \emph{inherited attributes}: values that are computed
top-down in the grammar. To the right of each symbol we have the
so-called \emph{synthesized attributes}: values that are computed
bottom-up. The arrows between attributes specify the information flow
to compute an attribute. For example, in the production \ensuremath{\Conid{Assign}}, to
compute the inherited attribute \ensuremath{\attrid{dcli}} in the third child (\ensuremath{\Conid{List}}), we
take the value of the first child (a \ensuremath{\bnfnt{Name}}) and insert it (with
\ensuremath{(\mathbin{:})}) to the beginning of the attribute \ensuremath{\attrid{dcli}} of the parent. That is
to say that we are accumulating identifiers in \ensuremath{\attrid{dcli}} while descending
in the list of statements of let expressions.


Thus, the AG expressed in Figure~\ref{fig:AG} is the following.
The inherited attribute \ensuremath{\attrid{dcli}} is used as an accumulator to
collect all \ensuremath{\Conid{Names}} defined in a \ensuremath{\Conid{Let}}: it starts as an empty list in
the \ensuremath{\Conid{Root}} production, and when a new name is defined (productions
\ensuremath{\Conid{Assign}} and \ensuremath{\Conid{NestedLet}}) it is added to the accumulator. The total
list of defined \ensuremath{\bnfnt{Name}} is synthesized in attribute \ensuremath{\attrid{dclo}}, which at
the \ensuremath{\Conid{Root}} node is passed down as the environment (inherited attribute
\ensuremath{\attrid{env}}). The type of the three attributes is a list of pairs,
associating the \ensuremath{\bnfnt{Name}} to its \ensuremath{\Conid{Let}} expression definition\footnote{We
will use this definition to expand the \ensuremath{\bnfnt{Name}} as required by
optimization rule \ensuremath{\mathrm{7}}.}.

Very much like strategic term rewriting, AGs also rely on a generic
tree walk mechanism, usually called tree-walk
evaluators~\cite{Alblas91b}, to walk up and down the tree to evaluate
attributes. In fact, generic zippers~\cite{thezipper} also offer the
necessary abstractions to express the embedding of AGs in a functional
programming setting~\cite{scp2016,memoAG19}. Next, we briefly describe
this embedding, and after that we present the embedded AG that express
the scope rules of \ensuremath{\Conid{Let}}. It also computes (attribute) \ensuremath{\attrid{env}}, that is
needed by the optimization rule $7$.

To allow programmers to write zipper-based functions very much
like attribute grammar writers do, the generic
zippers library~\cite{adams2010zippers} is extended with the following set of
simple AG-like combinators:

\begin{itemize}

\item The combinator ``\textit{child}'', written as the infix function
\ensuremath{\textsf{.\$}} to access the child of a tree node given its index (starting from
1).

\begin{hscode}\SaveRestoreHook
\column{B}{@{}>{\hspre}l<{\hspost}@{}}%
\column{E}{@{}>{\hspre}l<{\hspost}@{}}%
\>[B]{}(\textsf{.\$})\mathbin{::}\textbf{Zipper}\;\Varid{a}\to \Conid{Int}\to \textbf{Zipper}\;\Varid{a}{}\<[E]%
\ColumnHook
\end{hscode}\resethooks

\item The combinator \ensuremath{\textsf{parent}} to move the focus to the parent of a
tree node,

\begin{hscode}\SaveRestoreHook
\column{B}{@{}>{\hspre}l<{\hspost}@{}}%
\column{E}{@{}>{\hspre}l<{\hspost}@{}}%
\>[B]{}\textsf{parent}\mathbin{::}\textbf{Zipper}\;\Varid{a}\to \textbf{Zipper}\;\Varid{a}{}\<[E]%
\ColumnHook
\end{hscode}\resethooks

\item The combinators \ensuremath{\textsf{.\$\textless}} (left) and \ensuremath{\textsf{.\$\textgreater}} (right) navigate to the
i$^{th}$ sibling on the left/right of the current node:

\begin{hscode}\SaveRestoreHook
\column{B}{@{}>{\hspre}l<{\hspost}@{}}%
\column{E}{@{}>{\hspre}l<{\hspost}@{}}%
\>[B]{}(\textsf{.\$\textless}),(\textsf{.\$\textgreater})\mathbin{::}\textbf{Zipper}\;\Varid{a}\to \Conid{Int}\to \textbf{Zipper}\;\Varid{a}{}\<[E]%
\ColumnHook
\end{hscode}\resethooks

\end{itemize}

Having presented these zipper-based AG combinators, we can now express
the scope rules of \ensuremath{\Conid{Let}} expressions as an attribute grammar. To make
our definitions more AG friendly we define the following
(type) synonym:

\begin{hscode}\SaveRestoreHook
\column{B}{@{}>{\hspre}l<{\hspost}@{}}%
\column{16}{@{}>{\hspre}l<{\hspost}@{}}%
\column{E}{@{}>{\hspre}l<{\hspost}@{}}%
\>[B]{}\mathbf{type}\;\textsf{AGTree}\;\Varid{a}{}\<[16]%
\>[16]{}\mathrel{=}\textbf{Zipper}\;\Conid{Root}\to \Varid{a}{}\<[E]%
\ColumnHook
\end{hscode}\resethooks

We start by defining the equations of the synthesized attribute
\ensuremath{\attrid{dclo}}. For each definition of an occurrence of \ensuremath{\attrid{dclo}} we define an
equation in our zipper-based function. For example, in the diagrams of
the \ensuremath{\Conid{NestedLet}} and \ensuremath{\Conid{Assign}} productions we see that \ensuremath{\attrid{dclo}} is defined
as the \ensuremath{\attrid{dclo}} of the third child. Moreover, in production \ensuremath{\Conid{EmptyList}}
attribute \ensuremath{\attrid{dclo}} is a copy of \ensuremath{\attrid{dcli}}. This is exactly how such
equations are written in the zipper-based AG, as we can see in the
next function:

\begin{hscode}\SaveRestoreHook
\column{B}{@{}>{\hspre}l<{\hspost}@{}}%
\column{11}{@{}>{\hspre}l<{\hspost}@{}}%
\column{14}{@{}>{\hspre}l<{\hspost}@{}}%
\column{26}{@{}>{\hspre}l<{\hspost}@{}}%
\column{E}{@{}>{\hspre}l<{\hspost}@{}}%
\>[B]{}\attrid{dclo}\mathbin{::}\textsf{AGTree}\;[\mskip1.5mu (\bnfnt{Name},\textbf{Zipper}\;\Conid{Root})\mskip1.5mu]{}\<[E]%
\\
\>[B]{}\attrid{dclo}\;\Varid{t}\mathrel{=}{}\<[11]%
\>[11]{}\mathbf{case}\;(\textsf{constructor}\;\Varid{t})\;\mathbf{of}{}\<[E]%
\\
\>[11]{}\hsindent{3}{}\<[14]%
\>[14]{}\bnfprod{$\mathit{Let_{Let}}$}{}\<[26]%
\>[26]{}\to \attrid{dclo}\;(\Varid{t}\textsf{.\$}\mathrm{1}){}\<[E]%
\\
\>[11]{}\hsindent{3}{}\<[14]%
\>[14]{}\bnfprod{$\mathit{NestedLet_{List}}$}{}\<[26]%
\>[26]{}\to \attrid{dclo}\;(\Varid{t}\textsf{.\$}\mathrm{3}){}\<[E]%
\\
\>[11]{}\hsindent{3}{}\<[14]%
\>[14]{}\bnfprod{$\mathit{Assign_{List}}$}{}\<[26]%
\>[26]{}\to \attrid{dclo}\;(\Varid{t}\textsf{.\$}\mathrm{3}){}\<[E]%
\\
\>[11]{}\hsindent{3}{}\<[14]%
\>[14]{}\bnfprod{$\mathit{EmptyList_{List}}$}{}\<[26]%
\>[26]{}\to \attrid{dcli}\;\Varid{t}{}\<[E]%
\ColumnHook
\end{hscode}\resethooks

The function \ensuremath{\textsf{constructor}} and the constructors used in the case
alternatives is boilerplate code needed by the AG embedding. This code
has to be defined once per tree structure (\textit{i.e.}, AG), and can
be generated by using template \ensuremath{\Conid{Haskell}}~\cite{templatehaskell}.  To
provide a full implementation of our example, we include such code in
Appendix~\ref{sec:app}.

We may use the default rule of a case statement to express similar AG
equations. Consider the case of defining the inherited attribute
\ensuremath{\attrid{env}}. In most diagrams an occurrence of attribute \ensuremath{\attrid{env}} is defined as
a copy of the parent. There are two exceptions: in productions \ensuremath{\Conid{Root}}
and \ensuremath{\Conid{NestedLet}}. In both cases, \ensuremath{\attrid{env}} gets its value from the
synthesized attribute \ensuremath{\attrid{dclo}} of the same non-terminal/type. Thus, the
\ensuremath{\Conid{Haskell}} \ensuremath{\attrid{env}} function looks as follows:

\begin{hscode}\SaveRestoreHook
\column{B}{@{}>{\hspre}l<{\hspost}@{}}%
\column{10}{@{}>{\hspre}l<{\hspost}@{}}%
\column{11}{@{}>{\hspre}l<{\hspost}@{}}%
\column{18}{@{}>{\hspre}l<{\hspost}@{}}%
\column{E}{@{}>{\hspre}l<{\hspost}@{}}%
\>[B]{}\attrid{env}\mathbin{::}\textsf{AGTree}\;[\mskip1.5mu (\bnfnt{Name},\textbf{Zipper}\;\Conid{Root})\mskip1.5mu]{}\<[E]%
\\
\>[B]{}\attrid{env}\;\Varid{t}\mathrel{=}{}\<[10]%
\>[10]{}\mathbf{case}\;(\textsf{constructor}\;\Varid{t})\;\mathbf{of}{}\<[E]%
\\
\>[10]{}\hsindent{1}{}\<[11]%
\>[11]{}\bnfprod{$\mathit{Root_{P}}$}{}\<[18]%
\>[18]{}\to \attrid{dclo}\;\Varid{t}{}\<[E]%
\\
\>[10]{}\hsindent{1}{}\<[11]%
\>[11]{}\bnfprod{$\mathit{Let_{Let}}$}{}\<[18]%
\>[18]{}\to \attrid{dclo}\;\Varid{t}{}\<[E]%
\\
\>[10]{}\hsindent{1}{}\<[11]%
\>[11]{}\anonymous {}\<[18]%
\>[18]{}\to \attrid{env}\;(\textsf{parent}\;\Varid{t}){}\<[E]%
\ColumnHook
\end{hscode}\resethooks

Let us define now the accumulator attribute \ensuremath{\attrid{dcli}}.  The zipper
function when visiting nodes of type \ensuremath{\Conid{Let}}, has to consider two
alternatives: the parent node can be a \ensuremath{\Conid{Root}} or a \ensuremath{\Conid{NestedLet}} (the two
occurrences of \ensuremath{\Conid{Let}} as a child in the diagrams). This happens because
the rules to define its value differ: in the \ensuremath{\Conid{Root}} node it corresponds
to an empty list (our outermost \ensuremath{\Conid{Let}} is context-free), while in a
nested block, the accumulator \ensuremath{\attrid{dcli}} starts as the \ensuremath{\attrid{env}} of the outer
block. Thus, the zipper-based function \ensuremath{\attrid{dcli}} is expressed as follows:

\begin{hscode}\SaveRestoreHook
\column{B}{@{}>{\hspre}l<{\hspost}@{}}%
\column{7}{@{}>{\hspre}l<{\hspost}@{}}%
\column{8}{@{}>{\hspre}l<{\hspost}@{}}%
\column{14}{@{}>{\hspre}c<{\hspost}@{}}%
\column{14E}{@{}l@{}}%
\column{18}{@{}>{\hspre}l<{\hspost}@{}}%
\column{20}{@{}>{\hspre}l<{\hspost}@{}}%
\column{29}{@{}>{\hspre}c<{\hspost}@{}}%
\column{29E}{@{}l@{}}%
\column{32}{@{}>{\hspre}c<{\hspost}@{}}%
\column{32E}{@{}l@{}}%
\column{33}{@{}>{\hspre}l<{\hspost}@{}}%
\column{36}{@{}>{\hspre}l<{\hspost}@{}}%
\column{38}{@{}>{\hspre}l<{\hspost}@{}}%
\column{E}{@{}>{\hspre}l<{\hspost}@{}}%
\>[B]{}\attrid{dcli}\mathbin{::}\textsf{AGTree}\;[\mskip1.5mu (\bnfnt{Name},\textbf{Zipper}\;\Conid{Root})\mskip1.5mu]{}\<[E]%
\\
\>[B]{}\attrid{dcli}\;{}\<[7]%
\>[7]{}\Varid{t}\mathrel{=}\mathbf{case}\;(\textsf{constructor}\;\Varid{t})\;\mathbf{of}{}\<[E]%
\\
\>[7]{}\bnfprod{$\mathit{Root_{P}}$}{}\<[14]%
\>[14]{}\to {}\<[14E]%
\>[18]{}[\mskip1.5mu \mskip1.5mu]{}\<[E]%
\\
\>[7]{}\bnfprod{$\mathit{Let_{Let}}$}{}\<[14]%
\>[14]{}\to {}\<[14E]%
\>[18]{}\mathbf{case}\;(\textsf{constructor}\;(\textsf{parent}\;\Varid{t}))\;\mathbf{of}{}\<[E]%
\\
\>[18]{}\bnfprod{$\mathit{Root_{P}}$}{}\<[29]%
\>[29]{}\to {}\<[29E]%
\>[33]{}[\mskip1.5mu \mskip1.5mu]{}\<[E]%
\\
\>[18]{}\bnfprod{$\mathit{NestedLet_{List}}$}\to {}\<[33]%
\>[33]{}\attrid{env}\;{}\<[38]%
\>[38]{}(\textsf{parent}\;\Varid{t}){}\<[E]%
\\
\>[7]{}\hsindent{1}{}\<[8]%
\>[8]{}\anonymous {}\<[14]%
\>[14]{}\to {}\<[14E]%
\>[18]{}\mathbf{case}\;(\textsf{constructor}\;(\textsf{parent}\;\Varid{t}))\;\mathbf{of}{}\<[E]%
\\
\>[18]{}\hsindent{2}{}\<[20]%
\>[20]{}\bnfprod{$\mathit{Assign_{List}}$}{}\<[32]%
\>[32]{}\to {}\<[32E]%
\>[36]{}(\textsf{lexeme}\;(\textsf{parent}\;\Varid{t}),\textsf{parent}\;\Varid{t})\mathbin{:}(\attrid{dcli}\;(\textsf{parent}\;\Varid{t})){}\<[E]%
\\
\>[18]{}\hsindent{2}{}\<[20]%
\>[20]{}\bnfprod{$\mathit{NestedLet_{List}}$}{}\<[32]%
\>[32]{}\to {}\<[32E]%
\>[36]{}(\textsf{lexeme}\;(\textsf{parent}\;\Varid{t}),\textsf{parent}\;\Varid{t})\mathbin{:}(\attrid{dcli}\;(\textsf{parent}\;\Varid{t})){}\<[E]%
\\
\>[18]{}\hsindent{2}{}\<[20]%
\>[20]{}\bnfprod{$\mathit{Let_{Let}}$}{}\<[32]%
\>[32]{}\to {}\<[32E]%
\>[36]{}\attrid{dcli}\;(\textsf{parent}\;\Varid{t}){}\<[E]%
\ColumnHook
\end{hscode}\resethooks

In this AG function we use the function \ensuremath{\textsf{lexeme}}, which implements the
so-called \textit{syntactic references} in attribute
equations~\cite{Saraiva99}. In this case, \ensuremath{\textsf{lexeme}} returns the \ensuremath{\bnfnt{Name}}
argument of constructors \ensuremath{\Conid{Assign}} and \ensuremath{\Conid{NestedLet}}. This function is
other boilerplate code also included in appendix~\ref{sec:app}.

In order to specify the complete name analysis task of \ensuremath{\Conid{Let}}
expression we need to report which names violate the scope rules of
the language. In fact, we can modularly and incrementally extend our
zipper AG, and define a new (synthesized) attribute \ensuremath{\attrid{errors}} which
reports such violations. 
In Appendix~\ref{sec:app} we include its definition.
In the next section we will show how \ensuremath{\attrid{errors}} can be
expressed as a strategic function and combined with our AG.

\subsection{Strategic Attribute Grammars}
\label{sec:strategicAG}


By having embedding both strategic term rewriting and attribute
grammars in the same zipper-based setting,
and given that both are embedded as first-class citizens,
we can easily combine these
two powerful language engineering techniques. As a result, attribute
computations that do useful work on few productions/nodes can be
efficiently expressed via our \ensuremath{\Conid{Ztrategic}} library, while rewriting
rules that rely on context information can access attribute values.

\paragraph{Acessing Attribute Values from Strategies:} 

As we said in Section~\ref{sec3}, rule 7 of Figure~\ref{rules} cannot be implemented using a trivial strategy, since it depends on the context.
The rule states that a variable occurrence can be changed by its definition.
For this we need to compute an environment of definitions, which is what we have done with the attribute \ensuremath{\attrid{env}}, previously.
Thus, if we have access to such attribute in the definition of a strategy, we would be able to implement this rule.

Given that both attribute grammars and strategies use the zipper to walk through the tree, such combinations can be easily performed if
the strategy exposes the zipper, in order to be used to apply the given attribute.
This is done in our library by the \ensuremath{\Varid{adhocTPZ}} combinator:
\begin{hscode}\SaveRestoreHook
\column{B}{@{}>{\hspre}l<{\hspost}@{}}%
\column{3}{@{}>{\hspre}l<{\hspost}@{}}%
\column{13}{@{}>{\hspre}l<{\hspost}@{}}%
\column{E}{@{}>{\hspre}l<{\hspost}@{}}%
\>[3]{}\Varid{adhocTPZ}{}\<[13]%
\>[13]{}\mathbin{::}\Conid{TP}\;\Varid{e}\to (\Varid{a}\to \textbf{Zipper}\;\Varid{e}\to \Conid{Maybe}\;\Varid{b})\to \Conid{TP}\;\Varid{e}{}\<[E]%
\ColumnHook
\end{hscode}\resethooks
Notice that instead of taking a function of type \ensuremath{(\Varid{a}\to \Conid{Maybe}\;\Varid{b})}, as does the combinator \ensuremath{\Varid{adhocTP}} introduced in Section~\ref{sec2},
it receives a function of type \ensuremath{(\Varid{a}\to \textbf{Zipper}\;\Varid{e}\to \Conid{Maybe}\;\Varid{b})}, with the zipper as a parameter.

Then, we can define a function with this type, that implements rule 7:

\begin{hscode}\SaveRestoreHook
\column{B}{@{}>{\hspre}l<{\hspost}@{}}%
\column{17}{@{}>{\hspre}l<{\hspost}@{}}%
\column{20}{@{}>{\hspre}l<{\hspost}@{}}%
\column{22}{@{}>{\hspre}l<{\hspost}@{}}%
\column{31}{@{}>{\hspre}l<{\hspost}@{}}%
\column{E}{@{}>{\hspre}l<{\hspost}@{}}%
\>[B]{}\Varid{expC}\mathbin{::}\Conid{Exp}\to \textbf{Zipper}\;\Conid{Root}\to \Conid{Maybe}\;\Conid{Exp}{}\<[E]%
\\
\>[B]{}\Varid{expC}\;(\Conid{Var}\;\Varid{x})\;\Varid{z}{}\<[17]%
\>[17]{}\mathrel{=}{}\<[20]%
\>[20]{}\mathbf{case}\;\Varid{lookup}\;\Varid{x}\;(\attrid{env}\;\Varid{z})\;\mathbf{of}{}\<[E]%
\\
\>[20]{}\hsindent{2}{}\<[22]%
\>[22]{}\Conid{Just}\;\Varid{e}{}\<[31]%
\>[31]{}\to \textsf{lexeme\_Assign}\;\Varid{e}{}\<[E]%
\\
\>[20]{}\hsindent{2}{}\<[22]%
\>[22]{}\Conid{Nothing}{}\<[31]%
\>[31]{}\to \Conid{Nothing}{}\<[E]%
\\
\>[B]{}\Varid{expC}\;\anonymous \;\Varid{z}{}\<[17]%
\>[17]{}\mathrel{=}\Conid{Nothing}{}\<[E]%
\ColumnHook
\end{hscode}\resethooks
The variable \ensuremath{\Varid{x}} is searched in the environment returned by the \ensuremath{\attrid{env}} attribute;
in case it is found, the associated expression\footnote{The function \ensuremath{\textsf{lexeme\_Assign}} is another syntactic reference that in this case takes a \ensuremath{\textbf{Zipper}} and, if it is focused on an \ensuremath{\Conid{Assign}}, returns its expression.} is returned, otherwise the optimization is not performed.

Now we can combine this rule with the previously defined \ensuremath{\Varid{expr}}, that implements rules 1 to 6, and apply them to all nodes.

\begin{hscode}\SaveRestoreHook
\column{B}{@{}>{\hspre}l<{\hspost}@{}}%
\column{8}{@{}>{\hspre}l<{\hspost}@{}}%
\column{E}{@{}>{\hspre}l<{\hspost}@{}}%
\>[B]{}\Varid{opt''}{}\<[8]%
\>[8]{}\mathbin{::}\textbf{Zipper}\;\Conid{Root}\to \Conid{Maybe}\;(\textbf{Zipper}\;\Conid{Root}){}\<[E]%
\\
\>[B]{}\Varid{opt''}\;{}\<[8]%
\>[8]{}\Varid{r}\mathrel{=}\Varid{applyTP}\;(\Varid{innermost}\;\Varid{step})\;\Varid{r}{}\<[E]%
\\
\>[8]{}\mathbf{where}\;\Varid{step}\mathrel{=}\Varid{failTP}\mathbin{`\Varid{adhocTPZ}`}\Varid{expC}\mathbin{`\Varid{adhocTP}`}\Varid{expr}{}\<[E]%
\ColumnHook
\end{hscode}\resethooks

\paragraph{Synthesizing Attributes via Strategies:}

We have shown how attributes and strategies can be combined by using the former while defining the latter.
Now we show how to combine them the other way around; i.e. to express attribute computations as strategies.
As an example, let us consider the \ensuremath{\attrid{errors}} attribute, that returns
the list of names that violate the scope rules 
following the structure of input program.
We want to evaluate
the following input

\begin{minipage}[htb!]{.50\textwidth}
\begin{hscode}\SaveRestoreHook
\column{B}{@{}>{\hspre}l<{\hspost}@{}}%
\column{18}{@{}>{\hspre}l<{\hspost}@{}}%
\column{23}{@{}>{\hspre}l<{\hspost}@{}}%
\column{26}{@{}>{\hspre}l<{\hspost}@{}}%
\column{29}{@{}>{\hspre}l<{\hspost}@{}}%
\column{34}{@{}>{\hspre}l<{\hspost}@{}}%
\column{E}{@{}>{\hspre}l<{\hspost}@{}}%
\>[B]{}\Varid{letWithErrors}\mathrel{=}{}\<[18]%
\>[18]{}\mathbf{let}\;{}\<[23]%
\>[23]{}\Varid{a}{}\<[26]%
\>[26]{}\mathrel{=}\Varid{b}\mathbin{+}\mathrm{3}{}\<[E]%
\\
\>[23]{}\Varid{c}{}\<[26]%
\>[26]{}\mathrel{=}\mathrm{2}{}\<[E]%
\\
\>[23]{}\Varid{w}{}\<[26]%
\>[26]{}\mathrel{=}{}\<[29]%
\>[29]{}\mathbf{let}\;{}\<[34]%
\>[34]{}\Varid{c}\mathrel{=}\Varid{a}\mathbin{-}\Varid{b}{}\<[E]%
\\
\>[29]{}\mathbf{in}\;{}\<[34]%
\>[34]{}\Varid{c}\mathbin{+}\Varid{z}{}\<[E]%
\\
\>[23]{}\Varid{c}{}\<[26]%
\>[26]{}\mathrel{=}\Varid{c}\mathbin{+}\mathrm{3}\mathbin{-}\Varid{c}{}\<[E]%
\\
\>[18]{}\mathbf{in}\;{}\<[23]%
\>[23]{}(\Varid{a}\mathbin{+}\mathrm{7})\mathbin{+}\Varid{c}\mathbin{+}\Varid{w}{}\<[E]%
\ColumnHook
\end{hscode}\resethooks
\end{minipage}

to the list \ensuremath{[\mskip1.5mu \text{\ttfamily \char34 b\char34},\text{\ttfamily \char34 b\char34},\text{\ttfamily \char34 z\char34},\text{\ttfamily \char34 c\char34}\mskip1.5mu]}.
Recall that errors occur in two
situations: First, duplicated definitions that are efficiently
detected when a new \ensuremath{\bnfnt{Name}} (defined in nodes \ensuremath{\Conid{Assign}} and \ensuremath{\Conid{NestedLet}})
is accumulated in \ensuremath{\attrid{dcli}}. Then the newly defined \ensuremath{\bnfnt{Name}} \textit{must
not be in} the environment \ensuremath{\attrid{dcli}} accumulated till that
definition/node. This is expressed by the following zipper function:

\begin{hscode}\SaveRestoreHook
\column{B}{@{}>{\hspre}l<{\hspost}@{}}%
\column{21}{@{}>{\hspre}l<{\hspost}@{}}%
\column{31}{@{}>{\hspre}l<{\hspost}@{}}%
\column{54}{@{}>{\hspre}l<{\hspost}@{}}%
\column{E}{@{}>{\hspre}l<{\hspost}@{}}%
\>[B]{}\Varid{decls}\mathbin{::}\Conid{List}\to \textbf{Zipper}\;\Conid{Root}\to [\mskip1.5mu \bnfnt{Name}\mskip1.5mu]{}\<[E]%
\\
\>[B]{}\Varid{decls}\;(\Conid{Assign}\;{}\<[21]%
\>[21]{}\Varid{s}\;\anonymous \;\anonymous )\;\Varid{z}{}\<[31]%
\>[31]{}\mathrel{=}\Varid{mNBIn}\;(\textsf{lexeme}\;\Varid{z},\Varid{z})\;{}\<[54]%
\>[54]{}(\attrid{dcli}\;\Varid{z}){}\<[E]%
\\
\>[B]{}\Varid{decls}\;(\Conid{NestedLet}\;{}\<[21]%
\>[21]{}\Varid{s}\;\anonymous \;\anonymous )\;\Varid{z}{}\<[31]%
\>[31]{}\mathrel{=}\Varid{mNBIn}\;(\textsf{lexeme}\;\Varid{z},\Varid{z})\;{}\<[54]%
\>[54]{}(\attrid{dcli}\;\Varid{z}){}\<[E]%
\\
\>[B]{}\Varid{decls}\;\anonymous \;\anonymous {}\<[31]%
\>[31]{}\mathrel{=}[\mskip1.5mu \mskip1.5mu]{}\<[E]%
\ColumnHook
\end{hscode}\resethooks

Invalid uses are detected when a \ensuremath{\bnfnt{Name}} is used in an arithmetic
expression (\ensuremath{\Conid{Exp}}). In this case, the \ensuremath{\bnfnt{Name}} \textit{must be
in}\footnote{Functions \ensuremath{\Varid{mNBIn}} and \ensuremath{\Varid{mBIn}} are trivial lookup
functions. They are presented in Appendix~\ref{sec:app}.} the total
environment \ensuremath{\attrid{env}}.

\begin{hscode}\SaveRestoreHook
\column{B}{@{}>{\hspre}l<{\hspost}@{}}%
\column{14}{@{}>{\hspre}l<{\hspost}@{}}%
\column{17}{@{}>{\hspre}l<{\hspost}@{}}%
\column{E}{@{}>{\hspre}l<{\hspost}@{}}%
\>[B]{}\Varid{uses}\mathbin{::}\Conid{Exp}\to \textbf{Zipper}\;\Conid{Root}\to [\mskip1.5mu \bnfnt{Name}\mskip1.5mu]{}\<[E]%
\\
\>[B]{}\Varid{uses}\;(\Conid{Var}\;\Varid{i})\;\Varid{z}{}\<[17]%
\>[17]{}\mathrel{=}\Varid{mBIn}\;(\textsf{lexeme}\;\Varid{z})\;(\attrid{env}\;\Varid{z}){}\<[E]%
\\
\>[B]{}\Varid{uses}\;\anonymous \;{}\<[14]%
\>[14]{}\Varid{z}{}\<[17]%
\>[17]{}\mathrel{=}[\mskip1.5mu \mskip1.5mu]{}\<[E]%
\ColumnHook
\end{hscode}\resethooks

Now, we define a type-unifying strategy that produces as result the
list of errors. 

\begin{hscode}\SaveRestoreHook
\column{B}{@{}>{\hspre}l<{\hspost}@{}}%
\column{5}{@{}>{\hspre}l<{\hspost}@{}}%
\column{9}{@{}>{\hspre}l<{\hspost}@{}}%
\column{E}{@{}>{\hspre}l<{\hspost}@{}}%
\>[B]{}\attrid{errors}\mathbin{::}\textbf{Zipper}\;\Conid{Root}\to [\mskip1.5mu \bnfnt{Name}\mskip1.5mu]{}\<[E]%
\\
\>[B]{}\attrid{errors}\;{}\<[9]%
\>[9]{}\Varid{t}\mathrel{=}\Varid{applyTU}\;(\mathit{full\_tdTU}\;\Varid{step})\;\Varid{t}{}\<[E]%
\\
\>[B]{}\hsindent{5}{}\<[5]%
\>[5]{}\mathbf{where}\;\Varid{step}\mathrel{=}\Varid{failTU}\mathbin{`\Varid{adhocTUZ}`}\Varid{uses}\mathbin{`\Varid{adhocTUZ}`}\Varid{decls}{}\<[E]%
\ColumnHook
\end{hscode}\resethooks

Although the step function combines \ensuremath{\Varid{decls}} and \ensuremath{\Varid{uses}} in this order,
the resulting list does not report duplications first, and invalid
uses after. The strategic function \ensuremath{\Varid{adhocTUZ}} does combine the two
functions 
and the default failing function 
into one, which is applied while traversing (in a top-down
traversal) the tree. In fact, it produces the errors in the proper
order.

In Appendix~\ref{sec:app} we show the zipper-AG definition of \ensuremath{\attrid{errors}}, where most
of the attribute equations are just propagating attribute values
upwards without doing useful work! In fact, a type unifying strategy
provides a better solution for specifying such synthesized
computations since it focus on the productions/nodes where interesting
work has to be defined.  This is particularly relevant when we
consider the \ensuremath{\Conid{Let}} sub-language as part of a real programming language
(such as \ensuremath{\Conid{Haskell}} with its 116 constructors across 30 data
types). The difference in complexity between the strategic definition
of \ensuremath{\attrid{errors}} and its AG counterpart is much higher.

Attribute grammar systems provide the so-called attribute propagation
patterns to avoid polluting the specification with
\textit{copy-rules}. In most systems, a special notation and pre-fixed
behavior is associated with a set of off-the-shelf patterns that can
be reused across AGs~\cite{eli,uuag}. For example, in the UUAG
system~\cite{uuag}, the propagation patterns are the default rules
for any attribute. Thus, only the specific/interesting equations
have to be specified. However, being
a special notation, hard-wired to the AG system, makes the extension
or change of the existing rules almost impossible: the full system has
to be updated. Our embedding of strategic term rewriting provides a
powerful setting to express attribute propagation patterns: no special
additional notation/mechanism is needed.

\section{The \ensuremath{\Conid{Ztrategic}} Library}
\label{sec4}

As we have shown, zippers can express strategic term rewriting.  We 
developed a full library, named \ensuremath{\Conid{Ztrategic}}.
Figure~\ref{code:api} presents the full API, which is based on Strafunski's API,
but adding the possibility to manipulate the zipper that traverses the tree, in order to be able to, for example,
compute attributes.
We define two Strategy types, \ensuremath{\Conid{TP}} and \ensuremath{\Conid{TU}} which stand for Type-preserving and Type-Unifying. The former represents transformations, while the latter represents reductions. 

We define Primitive Strategies that can be used as building blocks, such as identities (\ensuremath{\Varid{idTP}} and \ensuremath{\Varid{constTU}}), failing strategies \ensuremath{\Varid{failTP}} and \ensuremath{\Varid{failTU}}, the \ensuremath{\Varid{tryTP}} combinator which applies a transformation once if possible and always succeeds, and \ensuremath{\Varid{repeatTP}} which applies a transformation as many times as possible. 
Strategy Construction combinators allow for composition of more complex transformations. The \ensuremath{\Varid{adhoc}} combinators compose an existing strategy with a \ensuremath{\Conid{Haskell}} function, and the \ensuremath{\Varid{mono}} combinators produce a strategy out of one \ensuremath{\Conid{Haskell}} function. There are variants that allow access to the zipper, denoted by a $Z$ suffix. 

For the composition of traversals, \ensuremath{\Varid{seq}} defines a sequence (perform both traversals) and \ensuremath{\Varid{choice}} an alternative (perform just one traversal). 
We use Traversal Combinators \ensuremath{\textbf{right}} and \ensuremath{\textbf{down}} to travel the zipper; the \ensuremath{\Varid{all}} variants always succeed and the \ensuremath{\Varid{one}} variants can fail. 
Traversal Strategies are defined by combining the previous tools, and come in \ensuremath{\Varid{td}} (top-down) and \ensuremath{\Varid{bu}} (bottom-up) variants. The \ensuremath{\Varid{full}} strategies traverse the whole tree, while \ensuremath{\Varid{once}} strategies perform at most one operation and \ensuremath{\Varid{stop}} strategies stop cut off the traversal in a specific sub-tree when any operation in it succeeds. Finally, \ensuremath{\Varid{innermost}} and \ensuremath{\Varid{outermost}} perform a transformation as many times as possible, starting from the inside or outside nodes, respectively.

\begin{figure*}[h]
\begin{minipage}[t]{.75\textwidth}
\textbf{Strategy types}\begin{hscode}\SaveRestoreHook
\column{B}{@{}>{\hspre}l<{\hspost}@{}}%
\column{3}{@{}>{\hspre}l<{\hspost}@{}}%
\column{E}{@{}>{\hspre}l<{\hspost}@{}}%
\>[3]{}\mathbf{type}\;\Conid{TP}\;\Varid{a}\mathrel{=}\textbf{Zipper}\;\Varid{a}\to \Conid{Maybe}\;(\textbf{Zipper}\;\Varid{a}){}\<[E]%
\\
\>[3]{}\mathbf{type}\;\Conid{TU}\;\Varid{m}\;\Varid{d}\mathrel{=}(\Varid{forall}\;\Varid{a}\;.\;\textbf{Zipper}\;\Varid{a}\to \Varid{m}\;\Varid{d}){}\<[E]%
\ColumnHook
\end{hscode}\resethooks
\textbf{Primitive strategies}\begin{hscode}\SaveRestoreHook
\column{B}{@{}>{\hspre}l<{\hspost}@{}}%
\column{3}{@{}>{\hspre}l<{\hspost}@{}}%
\column{13}{@{}>{\hspre}l<{\hspost}@{}}%
\column{E}{@{}>{\hspre}l<{\hspost}@{}}%
\>[3]{}\Varid{idTP}{}\<[13]%
\>[13]{}\mathbin{::}\Conid{TP}\;\Varid{a}{}\<[E]%
\\
\>[3]{}\Varid{constTU}{}\<[13]%
\>[13]{}\mathbin{::}\Varid{d}\to \Conid{TU}\;\Varid{m}\;\Varid{d}{}\<[E]%
\\
\>[3]{}\Varid{failTP}{}\<[13]%
\>[13]{}\mathbin{::}\Conid{TP}\;\Varid{a}{}\<[E]%
\\
\>[3]{}\Varid{failTU}{}\<[13]%
\>[13]{}\mathbin{::}\Conid{TU}\;\Varid{m}\;\Varid{d}{}\<[E]%
\\
\>[3]{}\Varid{tryTP}{}\<[13]%
\>[13]{}\mathbin{::}\Conid{TP}\;\Varid{a}\to \Conid{TP}\;\Varid{a}{}\<[E]%
\\
\>[3]{}\Varid{repeatTP}{}\<[13]%
\>[13]{}\mathbin{::}\Conid{TP}\;\Varid{a}\to \Conid{TP}\;\Varid{a}{}\<[E]%
\ColumnHook
\end{hscode}\resethooks
\textbf{Strategy Construction}\begin{hscode}\SaveRestoreHook
\column{B}{@{}>{\hspre}l<{\hspost}@{}}%
\column{3}{@{}>{\hspre}l<{\hspost}@{}}%
\column{13}{@{}>{\hspre}l<{\hspost}@{}}%
\column{E}{@{}>{\hspre}l<{\hspost}@{}}%
\>[3]{}\Varid{monoTP}{}\<[13]%
\>[13]{}\mathbin{::}(\Varid{a}\to \Conid{Maybe}\;\Varid{b})\to \Conid{TP}\;\Varid{e}{}\<[E]%
\\
\>[3]{}\Varid{monoTU}{}\<[13]%
\>[13]{}\mathbin{::}(\Varid{a}\to \Varid{m}\;\Varid{d})\to \Conid{TU}\;\Varid{m}\;\Varid{d}{}\<[E]%
\\
\>[3]{}\Varid{monoTPZ}{}\<[13]%
\>[13]{}\mathbin{::}(\Varid{a}\to \textbf{Zipper}\;\Varid{e}\to \Conid{Maybe}\;\Varid{b})\to \Conid{TP}\;\Varid{e}{}\<[E]%
\\
\>[3]{}\Varid{monoTUZ}{}\<[13]%
\>[13]{}\mathbin{::}(\Varid{a}\to \textbf{Zipper}\;\Varid{e}\to \Varid{m}\;\Varid{d})\to \Conid{TU}\;\Varid{m}\;\Varid{d}{}\<[E]%
\\
\>[3]{}\Varid{adhocTP}{}\<[13]%
\>[13]{}\mathbin{::}\Conid{TP}\;\Varid{e}\to (\Varid{a}\to \Conid{Maybe}\;\Varid{b})\to \Conid{TP}\;\Varid{e}{}\<[E]%
\\
\>[3]{}\Varid{adhocTU}{}\<[13]%
\>[13]{}\mathbin{::}\Conid{TU}\;\Varid{m}\;\Varid{d}\to (\Varid{a}\to \Varid{m}\;\Varid{d})\to \Conid{TU}\;\Varid{m}\;\Varid{d}{}\<[E]%
\\
\>[3]{}\Varid{adhocTPZ}{}\<[13]%
\>[13]{}\mathbin{::}\Conid{TP}\;\Varid{e}\to (\Varid{a}\to \textbf{Zipper}\;\Varid{e}\to \Conid{Maybe}\;\Varid{b})\to \Conid{TP}\;\Varid{e}{}\<[E]%
\\
\>[3]{}\Varid{adhocTUZ}{}\<[13]%
\>[13]{}\mathbin{::}\Conid{TU}\;\Varid{m}\;\Varid{d}\to (\Varid{a}\to \textbf{Zipper}\;\Varid{c}\to \Varid{m}\;\Varid{d})\to \Conid{TU}\;\Varid{m}\;\Varid{d}{}\<[E]%
\ColumnHook
\end{hscode}\resethooks
\textbf{Composition / Choice}\begin{hscode}\SaveRestoreHook
\column{B}{@{}>{\hspre}l<{\hspost}@{}}%
\column{3}{@{}>{\hspre}l<{\hspost}@{}}%
\column{13}{@{}>{\hspre}l<{\hspost}@{}}%
\column{E}{@{}>{\hspre}l<{\hspost}@{}}%
\>[3]{}\Varid{seqTP}{}\<[13]%
\>[13]{}\mathbin{::}\Conid{TP}\;\Varid{a}\to \Conid{TP}\;\Varid{a}\to \Conid{TP}\;\Varid{a}{}\<[E]%
\\
\>[3]{}\Varid{choiceTP}{}\<[13]%
\>[13]{}\mathbin{::}\Conid{TP}\;\Varid{a}\to \Conid{TP}\;\Varid{a}\to \Conid{TP}\;\Varid{a}{}\<[E]%
\\
\>[3]{}\Varid{seqTU}{}\<[13]%
\>[13]{}\mathbin{::}\Conid{TU}\;\Varid{m}\;\Varid{d}\to \Conid{TU}\;\Varid{m}\;\Varid{d}\to \Conid{TU}\;\Varid{m}\;\Varid{d}{}\<[E]%
\\
\>[3]{}\Varid{choiceTU}{}\<[13]%
\>[13]{}\mathbin{::}\Conid{TU}\;\Varid{m}\;\Varid{d}\to \Conid{TU}\;\Varid{m}\;\Varid{d}\to \Conid{TU}\;\Varid{m}\;\Varid{d}{}\<[E]%
\ColumnHook
\end{hscode}\resethooks
\end{minipage}
\begin{minipage}[t]{.20\textwidth}
\textbf{Traversal Combinators}\begin{hscode}\SaveRestoreHook
\column{B}{@{}>{\hspre}l<{\hspost}@{}}%
\column{3}{@{}>{\hspre}l<{\hspost}@{}}%
\column{15}{@{}>{\hspre}l<{\hspost}@{}}%
\column{E}{@{}>{\hspre}l<{\hspost}@{}}%
\>[3]{}\Varid{allTPright}{}\<[15]%
\>[15]{}\mathbin{::}\Conid{TP}\;\Varid{a}\to \Conid{TP}\;\Varid{a}{}\<[E]%
\\
\>[3]{}\Varid{oneTPright}{}\<[15]%
\>[15]{}\mathbin{::}\Conid{TP}\;\Varid{a}\to \Conid{TP}\;\Varid{a}{}\<[E]%
\\
\>[3]{}\Varid{allTUright}{}\<[15]%
\>[15]{}\mathbin{::}\Conid{TU}\;\Varid{m}\;\Varid{d}\to \Conid{TU}\;\Varid{m}\;\Varid{d}{}\<[E]%
\\
\>[3]{}\Varid{allTPdown}{}\<[15]%
\>[15]{}\mathbin{::}\Conid{TP}\;\Varid{a}\to \Conid{TP}\;\Varid{a}{}\<[E]%
\\
\>[3]{}\Varid{oneTPdown}{}\<[15]%
\>[15]{}\mathbin{::}\Conid{TP}\;\Varid{a}\to \Conid{TP}\;\Varid{a}{}\<[E]%
\\
\>[3]{}\Varid{allTUdown}{}\<[15]%
\>[15]{}\mathbin{::}\Conid{TU}\;\Varid{m}\;\Varid{d}\to \Conid{TU}\;\Varid{m}\;\Varid{d}{}\<[E]%
\ColumnHook
\end{hscode}\resethooks
\textbf{Traversal Strategies}\begin{hscode}\SaveRestoreHook
\column{B}{@{}>{\hspre}l<{\hspost}@{}}%
\column{3}{@{}>{\hspre}l<{\hspost}@{}}%
\column{14}{@{}>{\hspre}l<{\hspost}@{}}%
\column{E}{@{}>{\hspre}l<{\hspost}@{}}%
\>[3]{}\mathit{full\_tdTP}{}\<[14]%
\>[14]{}\mathbin{::}\Conid{TP}\;\Varid{a}\to \Conid{TP}\;\Varid{a}{}\<[E]%
\\
\>[3]{}\mathit{full\_buTP}{}\<[14]%
\>[14]{}\mathbin{::}\Conid{TP}\;\Varid{a}\to \Conid{TP}\;\Varid{a}{}\<[E]%
\\
\>[3]{}\mathit{once\_tdTP}{}\<[14]%
\>[14]{}\mathbin{::}\Conid{TP}\;\Varid{a}\to \Conid{TP}\;\Varid{a}{}\<[E]%
\\
\>[3]{}\mathit{once\_buTP}{}\<[14]%
\>[14]{}\mathbin{::}\Conid{TP}\;\Varid{a}\to \Conid{TP}\;\Varid{a}{}\<[E]%
\\
\>[3]{}\mathit{stop\_tdTP}{}\<[14]%
\>[14]{}\mathbin{::}\Conid{TP}\;\Varid{a}\to \Conid{TP}\;\Varid{a}{}\<[E]%
\\
\>[3]{}\mathit{stop\_buTP}{}\<[14]%
\>[14]{}\mathbin{::}\Conid{TP}\;\Varid{a}\to \Conid{TP}\;\Varid{a}{}\<[E]%
\\
\>[3]{}\Varid{innermost}{}\<[14]%
\>[14]{}\mathbin{::}\Conid{TP}\;\Varid{a}\to \Conid{TP}\;\Varid{a}{}\<[E]%
\\
\>[3]{}\Varid{outermost}{}\<[14]%
\>[14]{}\mathbin{::}\Conid{TP}\;\Varid{a}\to \Conid{TP}\;\Varid{a}{}\<[E]%
\\
\>[3]{}\mathit{full\_tdTU}{}\<[14]%
\>[14]{}\mathbin{::}\Conid{TU}\;\Varid{m}\;\Varid{d}\to \Conid{TU}\;\Varid{m}\;\Varid{d}{}\<[E]%
\\
\>[3]{}\mathit{full\_buTU}{}\<[14]%
\>[14]{}\mathbin{::}\Conid{TU}\;\Varid{m}\;\Varid{d}\to \Conid{TU}\;\Varid{m}\;\Varid{d}{}\<[E]%
\\
\>[3]{}\mathit{once\_tdTU}{}\<[14]%
\>[14]{}\mathbin{::}\Conid{TU}\;\Varid{m}\;\Varid{d}\to \Conid{TU}\;\Varid{m}\;\Varid{d}{}\<[E]%
\\
\>[3]{}\mathit{once\_buTU}{}\<[14]%
\>[14]{}\mathbin{::}\Conid{TU}\;\Varid{m}\;\Varid{d}\to \Conid{TU}\;\Varid{m}\;\Varid{d}{}\<[E]%
\\
\>[3]{}\mathit{stop\_tdTU}{}\<[14]%
\>[14]{}\mathbin{::}\Conid{TU}\;\Varid{m}\;\Varid{d}\to \Conid{TU}\;\Varid{m}\;\Varid{d}{}\<[E]%
\\
\>[3]{}\mathit{stop\_buTU}{}\<[14]%
\>[14]{}\mathbin{::}\Conid{TU}\;\Varid{m}\;\Varid{d}\to \Conid{TU}\;\Varid{m}\;\Varid{d}{}\<[E]%
\ColumnHook
\end{hscode}\resethooks
\end{minipage}
\caption{Full Ztrategic API}
\label{code:api}
\end{figure*}

Let us show now the expressiveness of our \ensuremath{\Conid{Ztrategic}} library in
implementing useful transformations of a real programming language
such as \ensuremath{\Conid{Haskell}}. We reuse the available support for parsing and
pretty printing as part of the \ensuremath{\Conid{Haskell}} core libraries (in the
\textit{haskell-src} package).

\subsection{Do-notation elimination}

We start by defining a refactoring that eliminates the syntactic sugar
introduced by the monadic \textit{do-notation}. In fact, we used this
notation in \ensuremath{\Varid{sumBZero'}}, and we can rewrite in an applicative
functional style as expressed by the monadic binding function \ensuremath{(\bind)}.

\begin{hscode}\SaveRestoreHook
\column{B}{@{}>{\hspre}l<{\hspost}@{}}%
\column{E}{@{}>{\hspre}l<{\hspost}@{}}%
\>[B]{}\Varid{sumBZero''}\mathbin{::}\Conid{Maybe}\;\Conid{Exp}{}\<[E]%
\\
\>[B]{}\Varid{sumBZero''}\mathrel{=}\textbf{down\textquoteright}\;\mathit{t_1}\bind\textbf{down\textquoteright}\bind\textbf{right}\bind\textbf{getHole}{}\<[E]%
\ColumnHook
\end{hscode}\resethooks

In order to automate this refactoring, a type-preserving strategy is
used to perform a full traversal in the \ensuremath{\Conid{Haskell}} tree, since such
expressions can be arbitrarily nested. The rewrite step
behaves like the identity function by default with a
type-specific case for pattern matching the \ensuremath{\mathbf{do}\mathbin{-}\Varid{notation}} in the
\ensuremath{\Conid{Haskell}} syntax tree (nodes constructed by \ensuremath{\Conid{HsDo}}).

\begin{hscode}\SaveRestoreHook
\column{B}{@{}>{\hspre}l<{\hspost}@{}}%
\column{5}{@{}>{\hspre}l<{\hspost}@{}}%
\column{11}{@{}>{\hspre}l<{\hspost}@{}}%
\column{E}{@{}>{\hspre}l<{\hspost}@{}}%
\>[B]{}\Varid{refactor}\mathbin{::}\textbf{Zipper}\;\Conid{HsModule}\to \Conid{Maybe}\;(\textbf{Zipper}\;\Conid{HsModule}){}\<[E]%
\\
\>[B]{}\Varid{refactor}\;{}\<[11]%
\>[11]{}\Varid{h}\mathrel{=}\Varid{applyTP}\;(\Varid{innermost}\;\Varid{step})\;\Varid{h}{}\<[E]%
\\
\>[B]{}\hsindent{5}{}\<[5]%
\>[5]{}\mathbf{where}\;\Varid{step}\mathrel{=}\Varid{failTP}\mathbin{`\Varid{adhocTP}`}\Varid{doElim}{}\<[E]%
\ColumnHook
\end{hscode}\resethooks

The following type-specific transformation function \ensuremath{\Varid{doElim}} just
expresses the refactoring we showed for the concrete \ensuremath{\Varid{sumBZero'}} to
\ensuremath{\Varid{sumBZero''}} example. Obviously, it is expressed at abstract syntax
tree level. We omit here the details of its underlying representation
as \ensuremath{\Conid{Haskell}} data types.

\begin{hscode}\SaveRestoreHook
\column{B}{@{}>{\hspre}l<{\hspost}@{}}%
\column{9}{@{}>{\hspre}l<{\hspost}@{}}%
\column{11}{@{}>{\hspre}l<{\hspost}@{}}%
\column{32}{@{}>{\hspre}l<{\hspost}@{}}%
\column{58}{@{}>{\hspre}l<{\hspost}@{}}%
\column{E}{@{}>{\hspre}l<{\hspost}@{}}%
\>[B]{}\Varid{doElim}{}\<[11]%
\>[11]{}\mathbin{::}\Conid{HsExp}\to \Conid{Maybe}\;\Conid{HsExp}{}\<[E]%
\\
\>[B]{}\Varid{doElim}\;(\Conid{HsDo}\;[\mskip1.5mu \Conid{HsQualifier}\;\Varid{e}\mskip1.5mu]){}\<[32]%
\>[32]{}\mathrel{=}\Conid{Just}\;\Varid{e}{}\<[E]%
\\
\>[B]{}\Varid{doElim}\;(\Conid{HsDo}\;(\Conid{HsQualifier}\;\Varid{e}\mathbin{:}\Varid{stmts}))\mathrel{=}\Conid{Just}\;((\Conid{HsInfixApp}\;{}\<[58]%
\>[58]{}\Varid{e}{}\<[E]%
\\
\>[B]{}\hsindent{9}{}\<[9]%
\>[9]{}(\Conid{HsQVarOp}\;(\Varid{hsSymbol}\;\text{\ttfamily \char34 >>\char34}))\;(\Conid{HsDo}\;\Varid{stmts}))){}\<[E]%
\\
\>[B]{}\Varid{doElim}\;(\Conid{HsDo}\;(\Conid{HsGenerator}\;\anonymous \;\Varid{p}\;\Varid{e}\mathbin{:}\Varid{stmts}))\mathrel{=}\Conid{Just}\;(\Varid{letPattern}\;\Varid{p}\;\Varid{e}\;\Varid{stmts}))\;{}\<[E]%
\\
\>[B]{}\Varid{doElim}\;(\Conid{HsDo}\;(\Conid{HsLetStmt}\;\Varid{decls}\mathbin{:}\Varid{stmts}))\mathrel{=}\Conid{Just}\;(\Conid{HsLet}\;\Varid{decls}\;(\Conid{HsDo}\;\Varid{stmts})){}\<[E]%
\\
\>[B]{}\Varid{doElim}\;\anonymous {}\<[11]%
\>[11]{}\mathrel{=}\Conid{Nothing}{}\<[E]%
\ColumnHook
\end{hscode}\resethooks

\subsection{Smells Elimination}

Source code smells make code harder to comprehend. A smell is not an
error, but it indicates a bad programming practice. They do occur in
any language and \ensuremath{\Conid{Haskell}} is no exception. For example, inexperienced
\ensuremath{\Conid{Haskell}} programmers often write \ensuremath{\Varid{l}\equiv [\mskip1.5mu \mskip1.5mu]} to check whether a list is
empty, instead of using the predefined \ensuremath{\Varid{null}} function.  Next, we
present a strategic function that eliminates several \ensuremath{\Conid{Haskell}} smells
as reported in~\cite{Cowie}.

\begin{hscode}\SaveRestoreHook
\column{B}{@{}>{\hspre}l<{\hspost}@{}}%
\column{3}{@{}>{\hspre}l<{\hspost}@{}}%
\column{12}{@{}>{\hspre}l<{\hspost}@{}}%
\column{17}{@{}>{\hspre}l<{\hspost}@{}}%
\column{25}{@{}>{\hspre}l<{\hspost}@{}}%
\column{52}{@{}>{\hspre}l<{\hspost}@{}}%
\column{E}{@{}>{\hspre}l<{\hspost}@{}}%
\>[B]{}\Varid{smellElim}{}\<[12]%
\>[12]{}\mathbin{::}\textbf{Zipper}\;\Conid{HsModule}\to \Conid{Maybe}\;(\textbf{Zipper}\;\Conid{HsModule}){}\<[E]%
\\
\>[B]{}\Varid{smellElim}\;\Varid{h}\mathrel{=}\Varid{applyTP}\;(\Varid{innermost}\;\Varid{step})\;\Varid{h}{}\<[E]%
\\
\>[B]{}\hsindent{3}{}\<[3]%
\>[3]{}\mathbf{where}\;\Varid{step}\mathrel{=}{}\<[17]%
\>[17]{}\Varid{failTP}{}\<[25]%
\>[25]{}\mathbin{`\Varid{adhocTP}`}\Varid{joinList}{}\<[52]%
\>[52]{}\mathbin{`\Varid{adhocTP}`}\Varid{nullList}{}\<[E]%
\\
\>[25]{}\mathbin{`\Varid{adhocTP}`}\Varid{redundantBoolean}\mathbin{`\Varid{adhocTP}`}\Varid{reduntantIf}{}\<[E]%
\ColumnHook
\end{hscode}\resethooks

where the type-specific transformations are described next. 

\begin{itemize}

\item \ensuremath{\Varid{joinList}} detects patterns where list concatenations are
inefficiently defined; the pattern \ensuremath{[\mskip1.5mu \Varid{x}\mskip1.5mu]\plus \Varid{xs}} is refactored to \ensuremath{\Varid{x}\mathbin{:}\Varid{xs}}.

\item \ensuremath{\Varid{nullList}} detects patterns where a list is checked for
emptiness. Patterns such as \ensuremath{\Varid{x}\equiv [\mskip1.5mu \mskip1.5mu]} and \ensuremath{\Varid{length}\;\Varid{x}\equiv \mathrm{0}} are
refactored to \ensuremath{\Varid{null}\;\Varid{x}}.

\item \ensuremath{\Varid{redundantBoolean}} detects redundant boolean checks, such as the
pattern \ensuremath{\Varid{x}\equiv \Conid{True}} which is refactored to \ensuremath{\Varid{x}}.

\item \ensuremath{\Varid{reduntantIf}} detects redundant usages of \ensuremath{\mathbf{if}} clauses, such as
the pattern \ensuremath{\mathbf{if}\;\Varid{x}\;\mathbf{then}\;\Conid{True}\;\mathbf{else}\;\Conid{False}} which is refactored to \ensuremath{\Varid{x}}.
\end{itemize}

The full implementation of these functions is included in
Appendix~\ref{sec:appSmell}.

\subsection{Ztrategic: Expressiveness and Performance}

In order to evaluate our combined zipper-based embedding of attribute
grammars and strategic term-rewriting we consider two language
engineering problems: First, we express in Ztrategic the largest
language specification developed in this setting: the Oberon-0
language. The construction of a processor for Oberon-0 was proposed in
the LDTA Tool Challenge~\cite{ldta}, and it was concisely and
efficiently specified using AGs and strategies in
Kiama~\cite{oberonkiama}. Second, we evaluate the performance of our
library by comparing the runtime of our Ztrategic smell eliminator
with its Strafunski counterpart when processing a large set of smelly
Haskell programs.

\paragraph{Oberon-0 in Ztrategic:}

The LDTA 2011 Tool Challenge~\cite{ldta} was a challenge focused on
the construction of a compiler for the Oberon-0 language, using the
tools of choice of the participants, with the goal of comparing the
formalisms and toolsets used in it. The challenge was divided into 5
tasks: parsing and pretty-printing, name binding, type checking,
desugaring and C code generation. These tasks were to be performed on
the Oberon-0 language, which in itself was divided into 5 increasingly
complex levels: (1) basic, (2) with if and while statements, (3) with
for and case statements, (4) with procedures and (5) with arrays and
records.

We consider the L2 level of the Oberon-0 language, and we specified
the name binding, type checking and desugaring tasks in our Ztrategic
AG approach. We use attributes for contextual information when needed,
for example in name analysis to check whether a used name has been
declared. This level requires the desugaring of \ensuremath{\Conid{For}} and \ensuremath{\Conid{Case}}
statements, into semantically equivalent \ensuremath{\Conid{While}} and (nested) \ensuremath{\Conid{If}}
statements. Such desugaring is implemented using Ztrategic type
preserving strategies, and the result is a (higher-order) tree that is
then decorated to perform name analysis and type checking via
ZippersAG. Because a \ensuremath{\Conid{For}} statement induces a new assignment (before
the \ensuremath{\Conid{WhileStmt}}) whose variable needs to be added to the declarations
part of the original AST, we use attribute \ensuremath{\Varid{numForDown}} which is a
synthesized attribute of the original tree.  Having the desugared AST
(represented by higher-order attributable attribute \ensuremath{\Varid{ata}}) and the
number of for statements refactored, then we return the final
higher-order tree where the induced variables are properly declared.

\begin{hscode}\SaveRestoreHook
\column{B}{@{}>{\hspre}l<{\hspost}@{}}%
\column{14}{@{}>{\hspre}l<{\hspost}@{}}%
\column{19}{@{}>{\hspre}l<{\hspost}@{}}%
\column{E}{@{}>{\hspre}l<{\hspost}@{}}%
\>[B]{}\Varid{desugar}\mathbin{::}\Conid{Root}\to \Conid{Root}{}\<[E]%
\\
\>[B]{}\Varid{desugar}\;\Varid{m}\mathrel{=}{}\<[14]%
\>[14]{}\mathbf{let}\;{}\<[19]%
\>[19]{}\Varid{numberOfFors}\mathrel{=}\Varid{numForsDown}\;(\textbf{toZipper}\;\Varid{m}){}\<[E]%
\\
\>[19]{}\Varid{step}\mathrel{=}\Varid{failTP}\mathbin{`\Varid{adhocTP}`}\Varid{desugarFor}\mathbin{`\Varid{adhocTPZ}`}\Varid{desugarCase}{}\<[E]%
\\
\>[19]{}\Varid{ata}\mathrel{=}\Varid{fromJust}\;(\Varid{applyTP}\;(\Varid{innermost}\;\Varid{step})\;(\textbf{toZipper}\;\Varid{m})){}\<[E]%
\\
\>[14]{}\mathbf{in}\;{}\<[19]%
\>[19]{}\Varid{injectForVars}\;\Varid{numberOfFors}\;(\Varid{fromZipper}\;\Varid{ata}){}\<[E]%
\ColumnHook
\end{hscode}\resethooks

Next, we show the work that is performed when the \ensuremath{\Varid{innermost}} strategy
visits \ensuremath{\Conid{ForStmt}} nodes:

\begin{hscode}\SaveRestoreHook
\column{B}{@{}>{\hspre}l<{\hspost}@{}}%
\column{9}{@{}>{\hspre}l<{\hspost}@{}}%
\column{13}{@{}>{\hspre}l<{\hspost}@{}}%
\column{23}{@{}>{\hspre}l<{\hspost}@{}}%
\column{31}{@{}>{\hspre}l<{\hspost}@{}}%
\column{39}{@{}>{\hspre}l<{\hspost}@{}}%
\column{E}{@{}>{\hspre}l<{\hspost}@{}}%
\>[B]{}\Varid{desugarFor}\mathbin{::}\Conid{Statement}\to \textbf{Zipper}\;\Conid{Root}\to \Conid{Maybe}\;\Conid{Statement}{}\<[E]%
\\
\>[B]{}\Varid{desugarFor}\;{}\<[13]%
\>[13]{}(\Conid{ForStmt}\;\Varid{idFor}\;\Varid{startFor}\;\Varid{dirFor}\;\Varid{stopFor}\;\Varid{stepFor}\;\Varid{ssFor})\;\Varid{z}\mathrel{=}{}\<[E]%
\\
\>[13]{}\Conid{Just}\;(\Conid{SeqStmt}\;\Varid{assign1}\;(\Conid{SeqStmt}\;\Varid{assign2}\;\Varid{loop})){}\<[E]%
\\
\>[B]{}\mathbf{where}\;{}\<[9]%
\>[9]{}\Varid{assign1}{}\<[23]%
\>[23]{}\mathrel{=}\Conid{AssigStmt}\;\Varid{idFor}\;\Varid{startFor}{}\<[E]%
\\
\>[9]{}\Varid{upperVarName}{}\<[23]%
\>[23]{}\mathrel{=}\text{\ttfamily \char34 \char95 forCounter\char34}\plus \Varid{show}\;(\Varid{numFors}\;\Varid{z}){}\<[E]%
\\
\>[9]{}\Varid{assign2}{}\<[23]%
\>[23]{}\mathrel{=}\Conid{AssigStmt}\;\Varid{upperVarName}\;\Varid{stopFor}{}\<[E]%
\\
\>[9]{}\Varid{loop}{}\<[23]%
\>[23]{}\mathrel{=}\Conid{WhileStmt}\;\Varid{whileExp}\;\Varid{whileStmt}{}\<[E]%
\\
\>[9]{}\Varid{whileExp}{}\<[23]%
\>[23]{}\mathrel{=}\Conid{IntCmpExp}\;\Varid{op}\;(\Conid{IdExp}\;\Varid{ifFor})\;(\Conid{IdExp}\;\Varid{upperVarName}){}\<[E]%
\\
\>[9]{}\Varid{op}{}\<[23]%
\>[23]{}\mathrel{=}\mathbf{case}\;{}\<[31]%
\>[31]{}\Varid{dirFor}\;\mathbf{of}{}\<[E]%
\\
\>[31]{}\Conid{To}{}\<[39]%
\>[39]{}\to \Conid{LECmp}{}\<[E]%
\\
\>[31]{}\Conid{Downto}{}\<[39]%
\>[39]{}\to \Conid{GECmp}{}\<[E]%
\\
\>[9]{}\Varid{whileStmt}{}\<[23]%
\>[23]{}\mathrel{=}\Conid{SeqStmt}\;\Varid{ssFor}\;\Varid{updateVar}{}\<[E]%
\\
\>[9]{}\Varid{updateVar}{}\<[23]%
\>[23]{}\mathrel{=}\Conid{AssigStmt}\;\Varid{ifFor}\;(\Conid{IntBOpExp}\;\Conid{Plus}\;(\Conid{IdExp}\;\Varid{ifFor})\;(\Varid{stepFor})){}\<[E]%
\\
\>[B]{}\Varid{desugarFor}\;\anonymous \;\Varid{z}\mathrel{=}\Conid{Nothing}{}\<[E]%
\ColumnHook
\end{hscode}\resethooks

\noindent
where, three new statements are produced: \ensuremath{\Varid{assign1}}, \ensuremath{\Varid{assign2}} and
\ensuremath{\Varid{loop}}, where the first two statements are assignments defining
variables for the initial and final values of the loop, respectively,
and \ensuremath{\Varid{loop}} is the transformed While loop. We also use attribute
\ensuremath{\Varid{numFors}} that counts the number of \ensuremath{\Conid{For}} loops that occur before this
loop, allowing us to identify this loop uniquely.

Next, we compare our approach to the results displayed
in~\cite{oberonkiama}, in terms of number of lines of code. For this,
we take their results for the L2 language level and compare it to
ours.

\begin{table}[]
\centering
\begin{tabular}{p{3cm} p{2cm} p{2cm}}
\hline
Task              & Ztrategic & Kiama \\ \hline
Oberon-0 Tree    & 57        & 99    \\
Name analyser     & 50        & 222   \\ 
Type analyser     & 34        & 117   \\ 
Lifter            & 6         & 23    \\
Desugarer         & 76        & 123   \\
Total             & 223       & 584   \\ \hline
\end{tabular}
\caption{Numbers of Lines of Code for the Oberon-0 L2 tasks.}
\end{table}

\paragraph{Ztrategic Smell Elimination:}

In order to assess the runtime performance of our zipper-based
strategic term rewriting implementation, we will compare it with the
state-of-the-art, fully optimized Strafunski system. A detailed
analysis of runtime performance of the zipper-based embedding of AGs
is presented in~\cite{memoAG19}. In this work we have also
incorporated memoization in order to avoid attribute recalculation. We
have executed several first and higher order zipper-based attribute
grammars with very large inputs, showing that our AG embedding is not
only concise and elegant, but is also efficient.

Let us consider the Haskell smell eliminator expressed both on
Ztrategic and Strafunski. To run both tools with large \textit{smelly}
inputs, we consider 150 Haskell projects developed by first-year
students as presented in~\cite{projects}. In these projects there are
1139 Haskell files totaling 82124 lines of code, of which exactly
1000 files were syntactically correct~\footnote{The student projects
used in this benchmark are available at this work's repository.}. Both
Ztrategic and Strafunski smell eliminators detected and eliminated 850
code smells in those files.

To compare the runtime performance of both implementations, we computed
an average of 5 runs, on a Ubuntu 16.04 machine, i5-7200U Dual Core,
with 8 GB RAM. 
In this case, the very first
version of Ztrategic is only 60\% slower than the Strafunski library:

\begin{table}[]
\centering
\begin{tabular}{p{4cm} p{2cm} p{2cm}}
\hline
                         & Ztrategic & Strafunski \\ \hline
Lines of Code            & 22        & 22         \\
Runtime                  & 16.2s     & 10.2s      \\
Average Memory Usage     & 6607.75Kb & 6580.16Kb  \\ \hline
\end{tabular}
\caption{Haskell Smell eliminators in Ztrategic and Strafunski.}
\end{table}

\section{Related Work}
\label{sec5}


The work we present in this paper is inspired by the pioneer work of
Sloane who developed Kiama~\cite{kiama}: an embedding of strategic
term rewriting and AGs in the Scala programming language.  In Kiama
attributes are defined as Scala functions, and strategic term
rewriting is expressed on Scala data structures via a set of strategic
combinators.
 While our
approach expresses both attribute computations and strategic term
rewriting as pure functions, Kiama caches attribute values in a global
cache, in order to reuse attribute values computed in the original
tree that are not affected by the rewriting. Such global cache,
however, induces an overhead in order to maintain it updated, that is,
attribute values associated to subtree discarded by the rewriting
process need to be purged from the cache~\cite{respectyourparents}. In
our purely functional setting, we only compute attributes in the
desired rewritten tree (as is the case of the let example shown in
section~\ref{sec3}).


Influenced by Kiama, Kramer and Van Wyk~\cite{strategicAG} present
\emph{strategy attributes}, which is an integration of strategic term
rewriting into attribute grammars. Strategic rewriting rules can use
the attributes of a tree to reference contextual information during
rewriting, much like we present in our work. Their work is
materialized as an extension to the extensible AG system
Silver~\cite{silver}.  They present several practical application
examples, namely the evaluation of $\lambda$-calculus, a regular
expression matching via Brzozowski derivatives, and the normalization
of for-loops. All these examples can be directly expressed in our
setting. They also present an application to optimize translation of
strategies. Because our techniques rely on shallow embeddings, where
no data type is used to express strategies nor AGs, we are not able to
express strategy optimizations, without relying on meta-programming
techniques~\cite{templatehaskell}. Nevertheless, our embeddings result
in very simple and small libraries that are easier to extend and
maintain, specially when compared to the complexity of extending and
maintaining a full language processor system such as Silver.


JastAdd is a reference attribute grammar based system~\cite{jastadd}. It
supports most of AG extensions, namely reference and circular
AGs~\cite{rag2013}. It also supports tree rewriting, with
rewrite rules that can reference attributes.  JastAdd, however,
provides no support for strategic programming, that is to say, there is
no mechanism to strategic control how the rewrite rules are applied.
The zipper-based AG embedding we integrate in \ensuremath{\Conid{Ztrategic}} supports all
modern AG extensions, including reference and circular
AGs~\cite{scp2016,memoAG19}. Because strategies and AGs are first
class citizens we can smoothly combine any of such extensions with
strategic term rewriting.


In the context of strategic term rewriting, our \ensuremath{\Conid{Ztrategic}} library
is inspired by Strafunski~\cite{strafunski}. In fact, \ensuremath{\Conid{Ztrategic}}
already provides almost all Strafunski functionality. There is,
however, a key difference between these libraries: while Strafunski
accesses the data structure directly, \ensuremath{\Conid{Ztrategic}} operates on
zippers. As a consequence, we can easily access attributes from
strategic functions and strategic functions from attribute equations.

\section{Conclusion}
\label{sec6}

This paper presented a zipper-based embedding of strategic term
rewriting. By relying on zippers, we combined it with a zipper-based
embedding of attribute grammars so that (zipper-based) strategies can
access (zipper-based) AG functional definitions, and vice versa. Thus,
zippers and strategies are first class citizens. We have developed the
\ensuremath{\Conid{Ztrategic}} strategic programming library and we have used
it to implement several language engineering tasks.

To assess the expressiveness of our approach we compared our Ztrategic
solution to the largest strategic AG developed by the state-of-the-art
Kiama system. In terms of runtime performance we compared our
Ztrategic library to the well established and fully optimized
Strafusnki solution. The preliminary results show that in fact zippers
provided a uniform setting to express both strategic term rewriting
and AGs that are in par with the state-of-the-art. Moreover, since
zippers do not rely on any advanced mechanism of our \ensuremath{\Conid{Haskell}} hosting
language, namely lazy evaluation, they can be implemented in other
(non-lazy) declarative programming languages. As a consequence, our
joint embeddings can easily be ported to any programming settings
where zippers are available.

\subsection{Future Work}

To avoid the re-calculation of attribute values, memoized zippers have
been incorporated in the embedding of the zipper-based
AG~\cite{memoAG19}. This results in a considerable performance
improvement when we run the embedded AG. Our strategic combinators can
also be expressed as memoized zippers, which will provide an
incremental setting for both strategic term rewriting and attribute
grammars. This will result in the rewriting of the equal sub-trees to
be performed once, only.

Our \ensuremath{\Conid{Ztrategic}} library can still be generalized to work with any \ensuremath{\Conid{Monad}\;\Varid{m}}
instead of being restricted to the \ensuremath{\Conid{Maybe}} monad. In fact. Strafunski
generalizes the monadic infrastructure used in their combinators, by
using, for example, the \ensuremath{\Conid{State}} monad to gather state-dependent
information, such as number of nodes traversed or number of failed
transformation applications. We will extend our library to incorporate
this generalization, too.

\bibliographystyle{splncs04}
\bibliography{bibliography}


\appendix

\section{The \ensuremath{\Conid{Let}} Attribute Grammar}
\label{sec:app}

The zipper-based AG function \ensuremath{\attrid{errors}} specifies the computation, via a
synthesized attribute, of the list of names that violate \ensuremath{\Conid{Let}} scope
rules. In Section~\ref{sec:strategicAG}, we have presented its
strategic counterpart definition that eliminates most of the copy (and gluing)
rules.

\begin{hscode}\SaveRestoreHook
\column{B}{@{}>{\hspre}l<{\hspost}@{}}%
\column{9}{@{}>{\hspre}l<{\hspost}@{}}%
\column{21}{@{}>{\hspre}l<{\hspost}@{}}%
\column{24}{@{}>{\hspre}l<{\hspost}@{}}%
\column{E}{@{}>{\hspre}l<{\hspost}@{}}%
\>[B]{}\attrid{errors}\mathbin{::}\textsf{AGTree}\;[\mskip1.5mu \Conid{String}\mskip1.5mu]{}\<[E]%
\\
\>[B]{}\attrid{errors}\;{}\<[9]%
\>[9]{}\Varid{t}\mathrel{=}\mathbf{case}\;(\textsf{constructor}\;\Varid{t})\;\mathbf{of}{}\<[E]%
\\
\>[9]{}\bnfprod{$\mathit{Root_{P}}$}{}\<[21]%
\>[21]{}\to \Varid{errs}\;(\Varid{t}\textsf{.\$}\mathrm{1}){}\<[E]%
\\
\>[9]{}\bnfprod{$\mathit{Let_{Let}}$}{}\<[21]%
\>[21]{}\to (\Varid{errs}\;(\Varid{t}\textsf{.\$}\mathrm{1}))\plus (\Varid{errs}\;(\Varid{t}\textsf{.\$}\mathrm{2})){}\<[E]%
\\
\>[9]{}\bnfprod{$\mathit{Add_{Exp}}$}{}\<[21]%
\>[21]{}\to (\Varid{errs}\;(\Varid{t}\textsf{.\$}\mathrm{1}))\plus (\Varid{errs}\;(\Varid{t}\textsf{.\$}\mathrm{2})){}\<[E]%
\\
\>[9]{}\bnfprod{$\mathit{Sub_{Exp}}$}{}\<[21]%
\>[21]{}\to (\Varid{errs}\;(\Varid{t}\textsf{.\$}\mathrm{1}))\plus (\Varid{errs}\;(\Varid{t}\textsf{.\$}\mathrm{2})){}\<[E]%
\\
\>[9]{}\bnfprod{$\mathit{EmptyList_{List}}$}{}\<[21]%
\>[21]{}\to [\mskip1.5mu \mskip1.5mu]{}\<[E]%
\\
\>[9]{}\bnfprod{$\mathit{Const_{Exp}}$}{}\<[21]%
\>[21]{}\to [\mskip1.5mu \mskip1.5mu]{}\<[E]%
\\
\>[9]{}\bnfprod{$\mathit{Var_{Exp}}$}{}\<[21]%
\>[21]{}\to \Varid{mBIn}\;(\textsf{lexeme}\;\Varid{t})\;(\attrid{env}\;\Varid{t}){}\<[E]%
\\
\>[9]{}\bnfprod{$\mathit{Assign_{List}}$}{}\<[21]%
\>[21]{}\to \Varid{mNBIn}\;(\textsf{lexeme}\;\Varid{t},\Varid{t})\;(\attrid{dcli}\;\Varid{t}){}\<[E]%
\\
\>[21]{}\hsindent{3}{}\<[24]%
\>[24]{}\plus (\Varid{errs}\;(\Varid{t}\textsf{.\$}\mathrm{2}))\plus (\Varid{errs}\;(\Varid{t}\textsf{.\$}\mathrm{3})){}\<[E]%
\\
\>[9]{}\bnfprod{$\mathit{NestedLet_{List}}$}{}\<[21]%
\>[21]{}\to \Varid{mNBIn}\;(\textsf{lexeme}\;\Varid{t},\Varid{t})\;(\attrid{dcli}\;\Varid{t}){}\<[E]%
\\
\>[21]{}\hsindent{3}{}\<[24]%
\>[24]{}\plus (\Varid{errs}\;(\Varid{t}\textsf{.\$}\mathrm{2}))\plus (\Varid{errs}\;(\Varid{t}\textsf{.\$}\mathrm{3})){}\<[E]%
\ColumnHook
\end{hscode}\resethooks

To distinguish the same name declared at different nesting levels, we
define an (inherited) attribute \ensuremath{\attrid{lev}}: the outermost let has level $0$
and we increment the level when descending to a let node. The next
zipper-based AG straightforwardly specifies the necessary attribute
equations.

\begin{hscode}\SaveRestoreHook
\column{B}{@{}>{\hspre}l<{\hspost}@{}}%
\column{10}{@{}>{\hspre}l<{\hspost}@{}}%
\column{17}{@{}>{\hspre}c<{\hspost}@{}}%
\column{17E}{@{}l@{}}%
\column{21}{@{}>{\hspre}l<{\hspost}@{}}%
\column{E}{@{}>{\hspre}l<{\hspost}@{}}%
\>[B]{}\attrid{lev}\mathbin{::}\textbf{Zipper}\;\Conid{Root}\to \Conid{Int}{}\<[E]%
\\
\>[B]{}\attrid{lev}\;\Varid{t}\mathrel{=}{}\<[10]%
\>[10]{}\mathbf{case}\;(\textsf{constructor}\;\Varid{t})\;\mathbf{of}{}\<[E]%
\\
\>[10]{}\bnfprod{$\mathit{Root_{P}}$}{}\<[17]%
\>[17]{}\to {}\<[17E]%
\>[21]{}\mathrm{0}{}\<[E]%
\\
\>[10]{}\bnfprod{$\mathit{Let_{Let}}$}{}\<[17]%
\>[17]{}\to {}\<[17E]%
\>[21]{}(\attrid{lev}\;(\textsf{parent}\;\Varid{t}))\mathbin{+}\mathrm{1}{}\<[E]%
\\
\>[10]{}\anonymous {}\<[17]%
\>[17]{}\to {}\<[17E]%
\>[21]{}\attrid{lev}\;(\textsf{parent}\;\Varid{t}){}\<[E]%
\ColumnHook
\end{hscode}\resethooks

Attribute grammars system usually provide a declarative language to
define auxiliary semantic functions. Functions \ensuremath{\Varid{mNBIn}} and \ensuremath{\Varid{mBIn}}
stand, respectively, for \textit{``must not be in"} and
\textit{``must be in"}, are simple lookup functions. Function \ensuremath{\Varid{mNBIn}}
needs the level to detect whether a name is erroneously defined at the
same level, or not. Obviously, we could store the level in the
environment. However, in our setting we can access attributes
decorating a tree in our semantic functions. Thus, these two functions
can be defined as follows:

\begin{hscode}\SaveRestoreHook
\column{B}{@{}>{\hspre}l<{\hspost}@{}}%
\column{23}{@{}>{\hspre}l<{\hspost}@{}}%
\column{39}{@{}>{\hspre}l<{\hspost}@{}}%
\column{E}{@{}>{\hspre}l<{\hspost}@{}}%
\>[B]{}\Varid{mBIn}\mathbin{::}\Conid{String}\to [\mskip1.5mu (\Conid{String},\textbf{Zipper}\;\Conid{Root})\mskip1.5mu]\to [\mskip1.5mu \Conid{String}\mskip1.5mu]{}\<[E]%
\\
\>[B]{}\Varid{mBIn}\;\Varid{name}\;[\mskip1.5mu \mskip1.5mu]{}\<[23]%
\>[23]{}\mathrel{=}[\mskip1.5mu \Varid{name}\mskip1.5mu]{}\<[E]%
\\
\>[B]{}\Varid{mBIn}\;\Varid{name}\;((\Varid{n},\Varid{l})\mathbin{:}\Varid{es}){}\<[23]%
\>[23]{}\mathrel{=}\mathbf{if}\;(\Varid{n}\equiv \Varid{name})\;{}\<[39]%
\>[39]{}\mathbf{then}\;[\mskip1.5mu \mskip1.5mu]{}\<[E]%
\\
\>[39]{}\mathbf{else}\;\Varid{mBIn}\;\Varid{name}\;\Varid{es}{}\<[E]%
\ColumnHook
\end{hscode}\resethooks

\begin{hscode}\SaveRestoreHook
\column{B}{@{}>{\hspre}l<{\hspost}@{}}%
\column{8}{@{}>{\hspre}l<{\hspost}@{}}%
\column{18}{@{}>{\hspre}c<{\hspost}@{}}%
\column{18E}{@{}l@{}}%
\column{21}{@{}>{\hspre}l<{\hspost}@{}}%
\column{E}{@{}>{\hspre}l<{\hspost}@{}}%
\>[B]{}\Varid{mNBIn}{}\<[8]%
\>[8]{}\mathbin{::}(\Conid{String},\textbf{Zipper}\;\Conid{Root})\to [\mskip1.5mu (\Conid{String},\textbf{Zipper}\;\Conid{Root})\mskip1.5mu]{}\<[E]%
\\
\>[8]{}\to [\mskip1.5mu \Conid{String}\mskip1.5mu]{}\<[E]%
\\
\>[B]{}\Varid{mNBIn}\;\Varid{tuple}\;[\mskip1.5mu \mskip1.5mu]{}\<[18]%
\>[18]{}\mathrel{=}{}\<[18E]%
\>[21]{}[\mskip1.5mu \mskip1.5mu]{}\<[E]%
\\
\>[B]{}\Varid{mNBIn}\;(\Varid{a1},\Varid{r1})\;((\Varid{a2},\Varid{r2})\mathbin{:}\Varid{es}){}\<[E]%
\\
\>[B]{}\hsindent{18}{}\<[18]%
\>[18]{}\mathrel{=}{}\<[18E]%
\>[21]{}\mathbf{if}\;(\Varid{a1}\equiv \Varid{a2})\mathrel{\wedge}(\attrid{lev}\;\Varid{r1}\equiv \attrid{lev}\;\Varid{r2}){}\<[E]%
\\
\>[21]{}\mathbf{then}\;[\mskip1.5mu \Varid{a1}\mskip1.5mu]\;\mathbf{else}\;\Varid{mNBIn}\;(\Varid{a1},\Varid{r1})\;\Varid{es}{}\<[E]%
\ColumnHook
\end{hscode}\resethooks

Zipper-based AG supports most modern extensions of attribute
grammars. For example, in~\cite{scp2016} these lookup functions are
expressed via higher-order attributes.  

\subsection{Boilerplate Code Induced by Data Types}

The code in this subsection is boilerplate code that needs to be built
once for each attribute grammar. Note that this code can be
automatically generated through Template
Haskell~\cite{templatehaskell}.

A \ensuremath{\Conid{Constructor}} data type is defined and it represents the data
constructor the zipper is focusing on and therefore must contain all
constructors in the data types.

\begin{hscode}\SaveRestoreHook
\column{B}{@{}>{\hspre}l<{\hspost}@{}}%
\column{19}{@{}>{\hspre}c<{\hspost}@{}}%
\column{19E}{@{}l@{}}%
\column{22}{@{}>{\hspre}l<{\hspost}@{}}%
\column{E}{@{}>{\hspre}l<{\hspost}@{}}%
\>[B]{}\mathbf{data}\;\Conid{Constructor}{}\<[19]%
\>[19]{}\mathrel{=}{}\<[19E]%
\>[22]{}\bnfprod{$\mathit{Root_{P}}$}{}\<[E]%
\\
\>[19]{}\mid {}\<[19E]%
\>[22]{}\bnfprod{$\mathit{Let_{Let}}$}{}\<[E]%
\\
\>[19]{}\mid {}\<[19E]%
\>[22]{}\bnfprod{$\mathit{NestedLet_{List}}$}{}\<[E]%
\\
\>[19]{}\mid {}\<[19E]%
\>[22]{}\bnfprod{$\mathit{Assign_{List}}$}{}\<[E]%
\\
\>[19]{}\mid {}\<[19E]%
\>[22]{}\bnfprod{$\mathit{EmptyList_{List}}$}{}\<[E]%
\\
\>[19]{}\mid {}\<[19E]%
\>[22]{}\bnfprod{$\mathit{Add_{Exp}}$}{}\<[E]%
\\
\>[19]{}\mid {}\<[19E]%
\>[22]{}\bnfprod{$\mathit{Sub_{Exp}}$}{}\<[E]%
\\
\>[19]{}\mid {}\<[19E]%
\>[22]{}\bnfprod{$\mathit{Neg_{Exp}}$}{}\<[E]%
\\
\>[19]{}\mid {}\<[19E]%
\>[22]{}\bnfprod{$\mathit{Const_{Exp}}$}{}\<[E]%
\\
\>[19]{}\mid {}\<[19E]%
\>[22]{}\bnfprod{$\mathit{Var_{Exp}}$}{}\<[E]%
\ColumnHook
\end{hscode}\resethooks

The \ensuremath{\textsf{constructor}} function looks at the zipper and returns the
aforementioned \ensuremath{\Conid{Constructor}} data value.

\begin{hscode}\SaveRestoreHook
\column{B}{@{}>{\hspre}l<{\hspost}@{}}%
\column{4}{@{}>{\hspre}l<{\hspost}@{}}%
\column{6}{@{}>{\hspre}l<{\hspost}@{}}%
\column{13}{@{}>{\hspre}l<{\hspost}@{}}%
\column{20}{@{}>{\hspre}l<{\hspost}@{}}%
\column{26}{@{}>{\hspre}l<{\hspost}@{}}%
\column{39}{@{}>{\hspre}c<{\hspost}@{}}%
\column{39E}{@{}l@{}}%
\column{42}{@{}>{\hspre}l<{\hspost}@{}}%
\column{44}{@{}>{\hspre}c<{\hspost}@{}}%
\column{44E}{@{}l@{}}%
\column{47}{@{}>{\hspre}l<{\hspost}@{}}%
\column{E}{@{}>{\hspre}l<{\hspost}@{}}%
\>[B]{}\textsf{constructor}\mathbin{::}\textbf{Zipper}\;\Varid{a}\to \Conid{Constructor}{}\<[E]%
\\
\>[B]{}\textsf{constructor}\;\Varid{ag}\mathrel{=}{}\<[E]%
\\
\>[B]{}\hsindent{4}{}\<[4]%
\>[4]{}\mathbf{case}\;(\textbf{getHole}\;\Varid{ag}\mathbin{::}\Conid{Maybe}\;\Conid{Root})\;\mathbf{of}{}\<[E]%
\\
\>[4]{}\hsindent{2}{}\<[6]%
\>[6]{}\Conid{Just}\;(\Conid{Root}\;\anonymous )\to \bnfprod{$\mathit{Root_{P}}$}{}\<[E]%
\\
\>[4]{}\hsindent{2}{}\<[6]%
\>[6]{}\anonymous \to \mathbf{case}\;(\textbf{getHole}\;\Varid{ag}\mathbin{::}\Conid{Maybe}\;\Conid{Let})\;\mathbf{of}{}\<[E]%
\\
\>[6]{}\hsindent{7}{}\<[13]%
\>[13]{}\Conid{Just}\;(\Conid{Let}\;\anonymous \;\anonymous )\to \bnfprod{$\mathit{Let_{Let}}$}{}\<[E]%
\\
\>[6]{}\hsindent{7}{}\<[13]%
\>[13]{}\anonymous \to \mathbf{case}\;(\textbf{getHole}\;\Varid{ag}\mathbin{::}\Conid{Maybe}\;\Conid{List})\;\mathbf{of}{}\<[E]%
\\
\>[13]{}\hsindent{7}{}\<[20]%
\>[20]{}\Conid{Just}\;(\Conid{NestedLet}\;\anonymous \;\anonymous \;\anonymous {}\<[44]%
\>[44]{}){}\<[44E]%
\>[47]{}\to \bnfprod{$\mathit{NestedLet_{List}}$}{}\<[E]%
\\
\>[13]{}\hsindent{7}{}\<[20]%
\>[20]{}\Conid{Just}\;(\Conid{Assign}\;\anonymous \;\anonymous \;\anonymous ){}\<[47]%
\>[47]{}\to \bnfprod{$\mathit{Assign_{List}}$}{}\<[E]%
\\
\>[13]{}\hsindent{7}{}\<[20]%
\>[20]{}\Conid{Just}\;(\Conid{EmptyList}{}\<[42]%
\>[42]{}){}\<[47]%
\>[47]{}\to \bnfprod{$\mathit{EmptyList_{List}}$}{}\<[E]%
\\
\>[13]{}\hsindent{7}{}\<[20]%
\>[20]{}\anonymous \to {}\<[26]%
\>[26]{}\mathbf{case}\;(\textbf{getHole}\;\Varid{ag}\mathbin{::}\Conid{Maybe}\;\Conid{Exp})\;\mathbf{of}{}\<[E]%
\\
\>[26]{}\Conid{Just}\;(\Conid{Add}\;\anonymous \;\anonymous ){}\<[42]%
\>[42]{}\to \bnfprod{$\mathit{Add_{Exp}}$}{}\<[E]%
\\
\>[26]{}\Conid{Just}\;(\Conid{Sub}\;\anonymous \;\anonymous ){}\<[42]%
\>[42]{}\to \bnfprod{$\mathit{Sub_{Exp}}$}{}\<[E]%
\\
\>[26]{}\Conid{Just}\;(\Conid{Neg}\;\anonymous {}\<[39]%
\>[39]{}){}\<[39E]%
\>[42]{}\to \bnfprod{$\mathit{Neg_{Exp}}$}{}\<[E]%
\\
\>[26]{}\Conid{Just}\;(\Conid{Var}\;\anonymous {}\<[39]%
\>[39]{}){}\<[39E]%
\>[42]{}\to \bnfprod{$\mathit{Var_{Exp}}$}{}\<[E]%
\\
\>[26]{}\Conid{Just}\;(\Conid{Const}\;\anonymous ){}\<[42]%
\>[42]{}\to \bnfprod{$\mathit{Const_{Exp}}$}{}\<[E]%
\\
\>[26]{}\anonymous {}\<[42]%
\>[42]{}\to \Varid{error}\;\text{\ttfamily \char34 Error\char34}{}\<[E]%
\ColumnHook
\end{hscode}\resethooks

The functions \ensuremath{\textsf{lexeme}} and \ensuremath{\textsf{lexeme\_Assign}} model syntactic references
in the AG, and they access information of certain nodes.

\begin{hscode}\SaveRestoreHook
\column{B}{@{}>{\hspre}l<{\hspost}@{}}%
\column{14}{@{}>{\hspre}l<{\hspost}@{}}%
\column{16}{@{}>{\hspre}l<{\hspost}@{}}%
\column{19}{@{}>{\hspre}l<{\hspost}@{}}%
\column{22}{@{}>{\hspre}l<{\hspost}@{}}%
\column{26}{@{}>{\hspre}l<{\hspost}@{}}%
\column{32}{@{}>{\hspre}l<{\hspost}@{}}%
\column{35}{@{}>{\hspre}l<{\hspost}@{}}%
\column{42}{@{}>{\hspre}l<{\hspost}@{}}%
\column{46}{@{}>{\hspre}l<{\hspost}@{}}%
\column{E}{@{}>{\hspre}l<{\hspost}@{}}%
\>[B]{}\textsf{lexeme}\mathbin{::}\textbf{Zipper}\;\Varid{a}\to \Conid{String}{}\<[E]%
\\
\>[B]{}\textsf{lexeme}\;\Varid{ag}\mathrel{=}{}\<[14]%
\>[14]{}\mathbf{case}\;(\textbf{getHole}\;\Varid{ag}\mathbin{::}\Conid{Maybe}\;\Conid{List})\;\mathbf{of}{}\<[E]%
\\
\>[14]{}\hsindent{5}{}\<[19]%
\>[19]{}\Conid{Just}\;(\Conid{Assign}\;{}\<[35]%
\>[35]{}\Varid{v}\;\anonymous \;\anonymous )\to \Varid{v}{}\<[E]%
\\
\>[14]{}\hsindent{5}{}\<[19]%
\>[19]{}\Conid{Just}\;(\Conid{NestedLet}\;\Varid{v}\;\anonymous \;\anonymous )\to \Varid{v}{}\<[E]%
\\
\>[14]{}\hsindent{5}{}\<[19]%
\>[19]{}\anonymous {}\<[22]%
\>[22]{}\to {}\<[26]%
\>[26]{}\mathbf{case}\;{}\<[32]%
\>[32]{}(\textbf{getHole}\;\Varid{ag}\mathbin{::}\Conid{Maybe}\;\Conid{Exp})\;\mathbf{of}{}\<[E]%
\\
\>[32]{}\Conid{Just}\;(\Conid{Var}\;\Varid{s}){}\<[46]%
\>[46]{}\to \Varid{s}{}\<[E]%
\\
\>[32]{}\anonymous {}\<[46]%
\>[46]{}\to \Varid{error}\;\text{\ttfamily \char34 Error\char34}{}\<[E]%
\\[\blanklineskip]%
\>[B]{}\textsf{lexeme\_Assign}{}\<[16]%
\>[16]{}\mathbin{::}\textbf{Zipper}\;\Varid{a}\to \Conid{Maybe}\;\Conid{Exp}{}\<[E]%
\\
\>[B]{}\textsf{lexeme\_Assign}\;{}\<[16]%
\>[16]{}\Varid{ag}\mathrel{=}{}\<[22]%
\>[22]{}\mathbf{case}\;(\textbf{getHole}\;\Varid{ag}\mathbin{::}\Conid{Maybe}\;\Conid{List})\;\mathbf{of}{}\<[E]%
\\
\>[22]{}\Conid{Just}\;(\Conid{Assign}\;\anonymous \;\Varid{e}\;\anonymous ){}\<[42]%
\>[42]{}\to \Conid{Just}\;\Varid{e}{}\<[E]%
\\
\>[22]{}\anonymous {}\<[42]%
\>[42]{}\to \Conid{Nothing}{}\<[E]%
\ColumnHook
\end{hscode}\resethooks

\section{Smell elimination}
\label{sec:appSmell}

We define refactoring \ensuremath{\Varid{smells}} through an \ensuremath{\Varid{innermost}} strategy that
applies a myriad of transformations, as many times as possible. The
use of \ensuremath{\Varid{innermost}} is necessary because performing one
refactoring can enable the application of another refactoring.

\begin{hscode}\SaveRestoreHook
\column{B}{@{}>{\hspre}l<{\hspost}@{}}%
\column{9}{@{}>{\hspre}l<{\hspost}@{}}%
\column{30}{@{}>{\hspre}l<{\hspost}@{}}%
\column{E}{@{}>{\hspre}l<{\hspost}@{}}%
\>[B]{}\Varid{smells}{}\<[9]%
\>[9]{}\mathbin{::}\textbf{Zipper}\;\Conid{HsModule}\to \Conid{Maybe}\;(\textbf{Zipper}\;\Conid{HsModule}){}\<[E]%
\\
\>[B]{}\Varid{smells}\;{}\<[9]%
\>[9]{}\Varid{h}\mathrel{=}\Varid{applyTP}\;(\Varid{innermost}\;\Varid{step})\;\Varid{h}{}\<[E]%
\\
\>[9]{}\mathbf{where}\;\Varid{step}\mathrel{=}\Varid{failTP}{}\<[30]%
\>[30]{}\mathbin{`\Varid{adhocTP}`}\Varid{joinList}{}\<[E]%
\\
\>[30]{}\mathbin{`\Varid{adhocTP}`}\Varid{nullList}{}\<[E]%
\\
\>[30]{}\mathbin{`\Varid{adhocTP}`}\Varid{redundantBoolean}{}\<[E]%
\\
\>[30]{}\mathbin{`\Varid{adhocTP}`}\Varid{reduntantIf}{}\<[E]%
\ColumnHook
\end{hscode}\resethooks

These functions are simple in the sense that they try to match a
pattern and replace them with another. They are similar in nature to
\ensuremath{\Varid{expr}} we defined in Section~\ref{sec2}, but the data types themselves
are more complex. We define a transformation to refactor the pattern
\ensuremath{[\mskip1.5mu \Varid{x}\mskip1.5mu]\plus \Varid{xs}} into \ensuremath{\Varid{x}\mathbin{:}\Varid{xs}}.

\begin{hscode}\SaveRestoreHook
\column{B}{@{}>{\hspre}l<{\hspost}@{}}%
\column{13}{@{}>{\hspre}l<{\hspost}@{}}%
\column{21}{@{}>{\hspre}l<{\hspost}@{}}%
\column{26}{@{}>{\hspre}l<{\hspost}@{}}%
\column{34}{@{}>{\hspre}l<{\hspost}@{}}%
\column{45}{@{}>{\hspre}l<{\hspost}@{}}%
\column{E}{@{}>{\hspre}l<{\hspost}@{}}%
\>[B]{}\Varid{joinList}\mathbin{::}\Conid{HsExp}\to \Conid{Maybe}\;\Conid{HsExp}{}\<[E]%
\\
\>[B]{}\Varid{joinList}\;{}\<[13]%
\>[13]{}(\Conid{HsInfixApp}\;{}\<[26]%
\>[26]{}(\Conid{HsList}\;[\mskip1.5mu \Varid{h}\mskip1.5mu])\;{}\<[E]%
\\
\>[26]{}(\Conid{HsQVarOp}\;(\Conid{UnQual}\;{}\<[45]%
\>[45]{}(\Conid{HsSymbol}\;\text{\ttfamily \char34 ++\char34})))\;{}\<[E]%
\\
\>[26]{}(\Conid{HsList}\;\Varid{t})){}\<[E]%
\\
\>[13]{}\mathrel{=}\Conid{Just}\;{}\<[21]%
\>[21]{}(\Conid{HsInfixApp}\;{}\<[34]%
\>[34]{}\Varid{h}\;{}\<[E]%
\\
\>[34]{}(\Conid{HsQConOp}\;(\Conid{Special}\;\Conid{HsCons}))\;{}\<[E]%
\\
\>[34]{}(\Conid{HsList}\;\Varid{t})){}\<[E]%
\\
\>[B]{}\Varid{joinList}\;\anonymous {}\<[13]%
\>[13]{}\mathrel{=}\Conid{Nothing}{}\<[E]%
\ColumnHook
\end{hscode}\resethooks

Next, we find patterns of bad null checking of lists, which are refactored to \ensuremath{\Varid{null}\;\Varid{x}}. The patterns are \ensuremath{\Varid{length}\;\Varid{x}\equiv \mathrm{0}}, \ensuremath{\mathrm{0}\equiv \Varid{length}\;\Varid{x}}, \ensuremath{\Varid{x}\equiv [\mskip1.5mu \mskip1.5mu]} and \ensuremath{[\mskip1.5mu \mskip1.5mu]\equiv \Varid{x}}, each represented by one line of the function.
\begin{hscode}\SaveRestoreHook
\column{B}{@{}>{\hspre}l<{\hspost}@{}}%
\column{3}{@{}>{\hspre}l<{\hspost}@{}}%
\column{11}{@{}>{\hspre}l<{\hspost}@{}}%
\column{13}{@{}>{\hspre}l<{\hspost}@{}}%
\column{23}{@{}>{\hspre}l<{\hspost}@{}}%
\column{24}{@{}>{\hspre}l<{\hspost}@{}}%
\column{31}{@{}>{\hspre}l<{\hspost}@{}}%
\column{37}{@{}>{\hspre}l<{\hspost}@{}}%
\column{38}{@{}>{\hspre}l<{\hspost}@{}}%
\column{E}{@{}>{\hspre}l<{\hspost}@{}}%
\>[B]{}\Varid{nullList}\mathbin{::}\Conid{HsExp}\to \Conid{Maybe}\;\Conid{HsExp}{}\<[E]%
\\
\>[B]{}\Varid{nullList}\;(\Conid{HsInfixApp}\;{}\<[23]%
\>[23]{}(\Conid{HsApp}\;{}\<[31]%
\>[31]{}(\Conid{HsVar}\;(\Conid{UnQual}{}\<[E]%
\\
\>[31]{}\hsindent{7}{}\<[38]%
\>[38]{}(\Conid{HsIdent}\;\text{\ttfamily \char34 length\char34})))\;{}\<[E]%
\\
\>[31]{}\Varid{a})\;{}\<[E]%
\\
\>[23]{}(\Conid{HsQVarOp}\;(\Conid{UnQual}\;(\Conid{HsSymbol}\;\text{\ttfamily \char34 ==\char34})))\;{}\<[E]%
\\
\>[23]{}(\Conid{HsLit}\;(\Conid{HsInt}\;\mathrm{0}))){}\<[E]%
\\
\>[B]{}\hsindent{3}{}\<[3]%
\>[3]{}\mathrel{=}\Conid{Just}\mathbin{\$}{}\<[13]%
\>[13]{}\Conid{HsApp}\;(\Conid{HsVar}\;(\Conid{UnQual}\;(\Conid{HsIdent}\;\text{\ttfamily \char34 null\char34})))\;\Varid{a}{}\<[E]%
\\
\>[B]{}\Varid{nullList}\;(\Conid{HsInfixApp}\;{}\<[23]%
\>[23]{}(\Conid{HsLit}\;(\Conid{HsInt}\;\mathrm{0}))\;{}\<[E]%
\\
\>[23]{}(\Conid{HsQVarOp}\;(\Conid{UnQual}\;(\Conid{HsSymbol}\;\text{\ttfamily \char34 ==\char34})))\;{}\<[E]%
\\
\>[23]{}(\Conid{HsApp}\;(\Conid{HsVar}\;(\Conid{UnQual}{}\<[E]%
\\
\>[23]{}\hsindent{14}{}\<[37]%
\>[37]{}(\Conid{HsIdent}\;\text{\ttfamily \char34 length\char34})))\;\Varid{a})){}\<[E]%
\\
\>[B]{}\hsindent{3}{}\<[3]%
\>[3]{}\mathrel{=}\Conid{Just}\mathbin{\$}\Conid{HsApp}\;(\Conid{HsVar}\;(\Conid{UnQual}\;(\Conid{HsIdent}\;\text{\ttfamily \char34 null\char34})))\;\Varid{a}{}\<[E]%
\\
\>[B]{}\Varid{nullList}\;{}\<[11]%
\>[11]{}(\Conid{HsInfixApp}\;{}\<[24]%
\>[24]{}\Varid{a}\;{}\<[E]%
\\
\>[24]{}(\Conid{HsQVarOp}\;(\Conid{UnQual}\;(\Conid{HsSymbol}\;\text{\ttfamily \char34 ==\char34})))\;{}\<[E]%
\\
\>[24]{}(\Conid{HsList}\;[\mskip1.5mu \mskip1.5mu])){}\<[E]%
\\
\>[B]{}\hsindent{3}{}\<[3]%
\>[3]{}\mathrel{=}\Conid{Just}\mathbin{\$}{}\<[13]%
\>[13]{}\Conid{HsApp}\;(\Conid{HsVar}\;(\Conid{UnQual}\;(\Conid{HsIdent}\;\text{\ttfamily \char34 null\char34})))\;\Varid{a}{}\<[E]%
\\
\>[B]{}\Varid{nullList}\;{}\<[11]%
\>[11]{}(\Conid{HsInfixApp}\;{}\<[24]%
\>[24]{}(\Conid{HsList}\;[\mskip1.5mu \mskip1.5mu])\;{}\<[E]%
\\
\>[24]{}(\Conid{HsQVarOp}\;(\Conid{UnQual}\;(\Conid{HsSymbol}\;\text{\ttfamily \char34 ==\char34})))\;{}\<[E]%
\\
\>[24]{}\Varid{a}){}\<[E]%
\\
\>[B]{}\hsindent{3}{}\<[3]%
\>[3]{}\mathrel{=}\Conid{Just}\mathbin{\$}{}\<[13]%
\>[13]{}\Conid{HsApp}\;(\Conid{HsVar}\;(\Conid{UnQual}\;(\Conid{HsIdent}\;\text{\ttfamily \char34 null\char34})))\;\Varid{a}{}\<[E]%
\\
\>[B]{}\Varid{nullList}\;\anonymous \mathrel{=}\Conid{Nothing}{}\<[E]%
\ColumnHook
\end{hscode}\resethooks

We remove redundant boolean checks, such as \ensuremath{\Varid{x}\equiv \Conid{True}} and \ensuremath{\Conid{True}\equiv \Varid{x}}, by refactoring them to \ensuremath{\Varid{x}}. Similarly, \ensuremath{\Varid{x}\equiv \Conid{False}} and \ensuremath{\Conid{False}\equiv \Varid{x}} are refactored to \ensuremath{\neg \;\Varid{x}}. 

\begin{hscode}\SaveRestoreHook
\column{B}{@{}>{\hspre}l<{\hspost}@{}}%
\column{3}{@{}>{\hspre}l<{\hspost}@{}}%
\column{19}{@{}>{\hspre}l<{\hspost}@{}}%
\column{E}{@{}>{\hspre}l<{\hspost}@{}}%
\>[B]{}\Varid{redundantBoolean}\mathbin{::}\Conid{HsExp}\to \Conid{Maybe}\;\Conid{HsExp}{}\<[E]%
\\
\>[B]{}\Varid{redundantBoolean}\;{}\<[19]%
\>[19]{}(\Conid{HsInfixApp}{}\<[E]%
\\
\>[19]{}(\Conid{HsCon}\;(\Conid{UnQual}\;(\Conid{HsIdent}\;\text{\ttfamily \char34 True\char34}))){}\<[E]%
\\
\>[19]{}(\Conid{HsQVarOp}\;(\Conid{UnQual}\;(\Conid{HsSymbol}\;\text{\ttfamily \char34 ==\char34}))){}\<[E]%
\\
\>[19]{}\Varid{a}){}\<[E]%
\\
\>[B]{}\hsindent{3}{}\<[3]%
\>[3]{}\mathrel{=}\Conid{Just}\;\Varid{a}{}\<[E]%
\\
\>[B]{}\Varid{redundantBoolean}\;{}\<[19]%
\>[19]{}(\Conid{HsInfixApp}{}\<[E]%
\\
\>[19]{}\Varid{a}{}\<[E]%
\\
\>[19]{}(\Conid{HsQVarOp}\;(\Conid{UnQual}\;(\Conid{HsSymbol}\;\text{\ttfamily \char34 ==\char34}))){}\<[E]%
\\
\>[19]{}(\Conid{HsCon}\;(\Conid{UnQual}\;(\Conid{HsIdent}\;\text{\ttfamily \char34 True\char34})))){}\<[E]%
\\
\>[B]{}\hsindent{3}{}\<[3]%
\>[3]{}\mathrel{=}\Conid{Just}\;\Varid{a}{}\<[E]%
\\
\>[B]{}\Varid{redundantBoolean}\;{}\<[19]%
\>[19]{}(\Conid{HsInfixApp}{}\<[E]%
\\
\>[19]{}(\Conid{HsCon}\;(\Conid{UnQual}\;(\Conid{HsIdent}\;\text{\ttfamily \char34 False\char34}))){}\<[E]%
\\
\>[19]{}(\Conid{HsQVarOp}\;(\Conid{UnQual}\;(\Conid{HsSymbol}\;\text{\ttfamily \char34 ==\char34}))){}\<[E]%
\\
\>[19]{}\Varid{a}){}\<[E]%
\\
\>[B]{}\hsindent{3}{}\<[3]%
\>[3]{}\mathrel{=}\Conid{Just}\mathbin{\$}(\Conid{HsApp}\;(\Conid{HsVar}\;(\Conid{UnQual}\;(\Conid{HsIdent}\;\text{\ttfamily \char34 not\char34})))\;\Varid{a}){}\<[E]%
\\
\>[B]{}\Varid{redundantBoolean}\;{}\<[19]%
\>[19]{}(\Conid{HsInfixApp}{}\<[E]%
\\
\>[19]{}\Varid{a}{}\<[E]%
\\
\>[19]{}(\Conid{HsQVarOp}\;(\Conid{UnQual}\;(\Conid{HsSymbol}\;\text{\ttfamily \char34 ==\char34}))){}\<[E]%
\\
\>[19]{}(\Conid{HsCon}\;(\Conid{UnQual}\;(\Conid{HsIdent}\;\text{\ttfamily \char34 False\char34})))){}\<[E]%
\\
\>[B]{}\hsindent{3}{}\<[3]%
\>[3]{}\mathrel{=}\Conid{Just}\mathbin{\$}(\Conid{HsApp}\;(\Conid{HsVar}\;(\Conid{UnQual}\;(\Conid{HsIdent}\;\text{\ttfamily \char34 not\char34})))\;\Varid{a}){}\<[E]%
\\
\>[B]{}\Varid{redundantBoolean}\;\anonymous \mathrel{=}\Conid{Nothing}{}\<[E]%
\ColumnHook
\end{hscode}\resethooks

Finally, we remove redundant usages of \ensuremath{\mathbf{if}} clauses by refactoring \ensuremath{\mathbf{if}\;\Varid{x}\;\mathbf{then}\;\Conid{True}\;\mathbf{else}\;\Conid{False}} into \ensuremath{\Varid{x}} and, conversely, \ensuremath{\mathbf{if}\;\Varid{x}\;\mathbf{then}\;\Conid{False}\;\mathbf{else}\;\Conid{True}} into \ensuremath{\neg \;\Varid{x}}. 

\begin{hscode}\SaveRestoreHook
\column{B}{@{}>{\hspre}l<{\hspost}@{}}%
\column{3}{@{}>{\hspre}l<{\hspost}@{}}%
\column{20}{@{}>{\hspre}l<{\hspost}@{}}%
\column{E}{@{}>{\hspre}l<{\hspost}@{}}%
\>[B]{}\Varid{reduntantIf}\mathbin{::}\Conid{HsExp}\to \Conid{Maybe}\;\Conid{HsExp}{}\<[E]%
\\
\>[B]{}\Varid{reduntantIf}\;(\Conid{HsIf}\;{}\<[20]%
\>[20]{}\Varid{a}\;{}\<[E]%
\\
\>[20]{}(\Conid{HsCon}\;(\Conid{UnQual}\;(\Conid{HsIdent}\;\text{\ttfamily \char34 True\char34})))\;{}\<[E]%
\\
\>[20]{}(\Conid{HsCon}\;(\Conid{UnQual}\;(\Conid{HsIdent}\;\text{\ttfamily \char34 False\char34})))){}\<[E]%
\\
\>[B]{}\hsindent{3}{}\<[3]%
\>[3]{}\mathrel{=}\Conid{Just}\;\Varid{a}{}\<[E]%
\\
\>[B]{}\Varid{reduntantIf}\;(\Conid{HsIf}\;{}\<[20]%
\>[20]{}\Varid{a}\;{}\<[E]%
\\
\>[20]{}(\Conid{HsCon}\;(\Conid{UnQual}\;(\Conid{HsIdent}\;\text{\ttfamily \char34 False\char34})))\;{}\<[E]%
\\
\>[20]{}(\Conid{HsCon}\;(\Conid{UnQual}\;(\Conid{HsIdent}\;\text{\ttfamily \char34 True\char34})))){}\<[E]%
\\
\>[B]{}\hsindent{3}{}\<[3]%
\>[3]{}\mathrel{=}\Conid{Just}\mathbin{\$}\Conid{HsApp}\;(\Conid{HsVar}\;(\Conid{UnQual}\;(\Conid{HsIdent}\;\text{\ttfamily \char34 not\char34})))\;\Varid{a}{}\<[E]%
\\
\>[B]{}\Varid{reduntantIf}\;\anonymous \mathrel{=}\Conid{Nothing}{}\<[E]%
\ColumnHook
\end{hscode}\resethooks

We can easily extend \ensuremath{\Varid{smells}} by implementing more refactor
transformations as simple \ensuremath{\Conid{Haskell}} functions and appending them in
the definition of \ensuremath{\Varid{step}}.

\end{document}